\documentclass[final,10pt,square,preprint,times]{elsarticle}
\usepackage{setspace}
\usepackage{geometry}
\usepackage{amssymb}
\usepackage{array}
\usepackage{booktabs}
\usepackage{tabularx}
\usepackage{makecell}
\usepackage{mathtools} % amsmath with fixes and additions
\usepackage{tikz} % nice language for creating drawings and diagrams
\usepackage{multirow}
\usepackage{algorithm}
\usepackage{algpseudocode}
\usepackage{amsmath}
\usepackage{array,graphicx}
\usepackage[colorlinks,linkcolor=black]{hyperref}
\usepackage[farskip=0pt]{subfig}
\newcommand{\tabincell}[2]{\begin{tabular}{@{}#1@{}}#2\end{tabular}}
\usepackage{array}
\usepackage{booktabs}
\journal{Journal of Speech Communication}

\begin{document}
\begin{frontmatter}

%% Title, authors and addresses

%% use the tnoteref command within \title for footnotes;
%% use the tnotetext command for theassociated footnote;
%% use the fnref command within \author or \affiliation for footnotes;
%% use the fntext command for theassociated footnote;
%% use the corref command within \author for corresponding author footnotes;
%% use the cortext command for theassociated footnote;
%% use the ead command for the email address,
%% and the form \ead[url] for the home page:
%% \title{Title\tnoteref{label1}}
%% \tnotetext[label1]{}
%% \author{Name\corref{cor1}\fnref{label2}}
%% \ead{email address}
%% \ead[url]{home page}
%% \fntext[label2]{}
%% \cortext[cor1]{}
%% \affiliation{organization={},
%%            addressline={}, 
%%            city={},
%%            postcode={}, 
%%            state={},
%%            country={}}
%% \fntext[label3]{}

\title{GM-TCNet: Gated Multi-scale Temporal Convolutional Network using Emotion Causality for Speech Emotion Recognition}
%% use optional labels to link authors explicitly to addresses:
%% \author[label1,label2]{}
%% \affiliation[label1]{organization={},
%%             addressline={},
%%             city={},
%%             postcode={},
%%             state={},
%%             country={}}
%%
%% \affiliation[label2]{organization={},
%%             addressline={},
%%             city={},
%%             postcode={},
%%             state={},
%%             country={}}
\author[1,2]{Jia-Xin Ye\fnref{fn1}}
\ead{jxye22@m.fudan.edu.cn}
\author[1,3]{Xin-Cheng Wen\fnref{fn1}}
\ead{xiamenwxc@foxmail.com}
\author[1]{Xuan-Ze Wang}
\ead{xuanze.wang@foxmail.com}
\author[4]{Yong Xu}
\ead{y.xu@fjut.edu.cn}
\author[1]{Yan Luo}
\ead{lauren-ly@foxmail.com}
\author[1]{Chang-Li Wu}
\ead{changli-wu@foxmail.com}
\author[5]{Li-Yan Chen}
\ead{chenliyan@xmu.edu.cn}
\author[1,5]{Kun-Hong Liu\corref{cor1}}
\ead{lkhqz@xmu.edu.cn}

\address[1]{School of Informatics, Xiamen University, Xiamen, China}
\address[2]{Institute of Science and Technology for Brain-inspired Intelligence, Fudan University, Shanghai, China}
\address[3]{Department of Computer Science, Harbin Institute of Technology (Shenzhen), Shenzhen, China}
\address[4]{School of Computer Science and Mathematics, Fujian University of Technology, Fuzhou, China}
\address[5]{School of Film, Xiamen University, Xiamen, China}

\fntext[fn1]{These authors contributed equally to this work.}
\cortext[cor1]{Corresponding author.}

\begin{abstract}
%% Text of abstract
In human-computer interaction, Speech Emotion Recognition (SER) plays an essential role in understanding the user's intent and improving the interactive experience. While similar sentimental speeches own diverse speaker characteristics but share common antecedents and consequences, an essential challenge for SER is how to produce robust and discriminative representations through causality between speech emotions. In this paper, we propose a Gated Multi-scale Temporal Convolutional Network (GM-TCNet) to construct a novel emotional causality representation learning component with a multi-scale receptive field. GM-TCNet deploys a novel emotional causality representation learning component to capture the dynamics of emotion across the time domain, constructed with dilated causal convolution layer and gating mechanism. Besides, it utilizes skip connection fusing high-level features from different gated convolution blocks to capture abundant and subtle emotion changes in human speech. GM-TCNet first uses a single type of feature, mel-frequency cepstral coefficients, as inputs and then passes them through the gated temporal convolutional module to generate the high-level features. Finally, the features are fed to the emotion classifier to accomplish the SER task. The experimental results show that our model maintains the highest performance in most cases compared to state-of-the-art techniques. The source code is available at: \href{https://github.com/Jiaxin-Ye/GM-TCNet_for_SER}{https://github.com/Jiaxin-Ye/GM-TCNet}.
\end{abstract}

%%Research highlights
% \begin{highlights}
% \item This paper proposes a novel network architecture called GM-TCNet for Speech Emotion Recognition based on the dilated causal convolution and gating mechanism.
% \item A novel emotional causality representation learning component is designed to capture the dynamics of emotion across the temporal domain, and better model the speech emotions at the frame level. It also has a strong ability in building a reliable long-term sentimental dependency. To the best of our knowledge, this is the first attempt at applying the emotional causality method to SER.
% \item GM-TCNet uses the skip connection among all Gated Convolution Blocks. It provides our network structure with a multi-scale temporal receptive field to improve its generalization ability. Moreover, a new dilated rate distribution of blocks is designed to obtain a larger receptive field, better fitting the SER applications.
% \item The proposed GM-TCNet approach gains state-of-the-art results in four widely studied datasets compared with other advanced approaches.
% \end{highlights}
\end{frontmatter}
% \footnotetext[1]{These authors contributed equally to this work.}

\section{Introduction}

Human-computer interaction (HCI) involves the study of the design and usage of computer technologies. It focuses not only on creating a natural and effective environment for interaction between humans and computers but also on providing a friendly interactive experience for users. As human speech signals are abundant of information, they convey the intent of messages through factors such as the speaker's identity, emotion, and intonation \cite{ser_contain_i}. For HCI applications, the identification of these factors behind the speech signals, especially the emotion, can enhance the understanding of the user's intent and improve the experience during the interaction. Therefore, the speech emotion recognition (SER) task that empowers machines to perceive emotion in human speech is becoming increasingly prevalent in the HCI field \cite{intro00}. SER develops rapidly and has been applied to many fields recently. For example, it had been widely used in the HCI \cite{intro03,intro04}, and Chen \textit{et al.} \cite{sota7} had embedded the SER technique in robots, so that robots could track a variety of emotions instantly. Moreover, it was also used to determine the suicidal tendency of patients \cite{intro01} and to prevent drivers from traffic accidents \cite{intro02}.

The typical SER system mainly includes two parts: feature extraction and emotion classification. For speech emotion feature extraction, it is mainly divided into qualitative features, spectral features, continuous features, and manually extracted features \cite{domain03,domain04}. Nowadays, spectral features have been extensively used, such as the Mel-Frequency Cepstral Coefficients (MFCC) \cite{domain01}, Linear Predictor Coefficients (LPC) \cite{domain02}. Besides, some studies \cite{sota24} tried to combine MFCC, Linear Predictive Cepstral Coefficient (LPCC) \cite{intro_add1} and other spectral features to tackle the SER task. At the same time, more and more manually extracted features are also introduced. For example, Tuncer \textit{et al.} \cite{sota17} selected the features by the shuffle box for feature generation and Mustaqeem \textit{et al.} \cite{sota12} used Radial Based Function Network (RBFN) similarity measurement to select a key sequence segment. These methods achieved high performance on multiple datasets by generating informative features in diversified ways.

Recently, researchers have proposed many different Deep Learning (DL) methods for the SER task, which can be divided into three categories: Generative Deep Learning (GDL), Discriminative Deep Learning (DDL) and Hybrid Deep Learning (HDL). Some typical models of the GDLs include the Generative Adversarial Nets (GAN), Deep Restricted Boltzmann Machine (DRBM) and Deep Auto-Encoders (DAE). Fang \textit{et al.} \cite{gan_ser} proposed a CycleGAN-based method to transfer features extracted from a large unlabeled speech corpus to synthetic features to represent the given target emotions. Zhang \textit{et al.} \cite{intro1} used the DRBM to learn relations between high-dimensional features and Fei \textit{et al.} \cite{intro2} utilized DAE to extract the features from raw speech. However, GDLs suffer from the vanishing gradient problem due to the sigmoid cross-entropy loss function used for training \cite{intro3}.

DDL has been more extensively used in SER compared to GDL, such as Convolutional Neural Network (CNN) \cite{intro4,intro5,intro6} and Recurrent Neural Network (RNN) \cite{intro06}. These techniques combined different layers in networks to provide high discriminative ability and eliminate the dependence on expert-driven handcrafted features. For example, to obtain a simpler and more general classification model, Issa \textit{et al.} \cite{sota13} adopted the 1-D CNN and stacked different speech features as input. Kwon \textit{et al.} \cite{intro11} proposed a model based on a 1-D dilated CNN (DCNN) with a multi-learning strategy to learn spatial and temporal features parallelly. Zhang \textit{et al.} \cite{intro15} designed a model based on multiple deep CNNs, which comprised 1-D, 2-D, and 3-D CNN to integrate different utterance-level results. Owing to the strong extensibility of CNN, researchers began to inject other networks into the CNN structure \cite{intro12}. However, CNN cannot effectively model temporal dependencies in the series data \cite{intro05}. Therefore, other DL frameworks with significant capability in handling sequential data, such as RNN, had been adopted in SER to preserve the temporality of speech signals~\cite{RNN_temporal}. For instance, Xie \textit{et al.}\cite{intro7} proposed the Long Short-Term Memory (LSTM) based on attention and gating mechanisms to control information flow by point-wise multiplication to capture dependencies from sequences and regulate the information at each frame. Su \textit{et al.}\cite{relate5} proposed a Graph Attention mechanism on the Gated Recurrent Unit network (GA-GRU) to handle the SER task, and Lin \textit{et al.}\cite{relate6} combined gated network and LSTM networks in a flexible way to preserve the temporal information of the sentence.

Furthermore, HDLs have been widely used nowadays because they inherit the advantages of different neural network structures. Zhao \textit{et al.}\cite{intro9} introduced the 1-D \& 2-D CNN-LSTM networks consisting of four Local Feature Learning Blocks (LFLB) and one LSTM layer. Each LFLB mainly contains one convolutional layer and one max-pooling layer to extract hierarchical correlations, and the LSTM layer learns the long-term dependencies from the local features. Meng \textit{et al.}\cite{intro10} introduced a fusion model comprising dilated CNN with residual block and Bidirectional Long-Short Term Memory (BiLSTM) based on the attention mechanism. Kwon \textit{et al.}\cite{intro13} utilized the hierarchical blocks of the Convolutional LSTM (ConvLSTM) and designed a new LFLB to capture the local emotional features in a hierarchical correlation. Furthermore, Temporal Convolutional Neural Network (TCNN) \cite{TCN1}, a modification based on CNN, was likewise used to maintain the temporal information of speech signals. It offers the capability of large-scale parallel processing with low training costs because it does not process the sequence data sequentially like RNN~\cite{TCN1}, so as to avoid high training costs~\cite{RNN_cost}.

%The speech signals take the form of the 2-D vector in the time domain and frequency domain. It is believed that the temporal information of the speech sequence is one of the indispensable pieces in the process of SER, which should be retained and investigated. Recently, CNNs have demonstrated superior performance in terms of SER. However, the CNN architecture can not capture the core temporal relationships across the time domain \cite{CNN_temporal}. Compared with CNN, RNN preserves the temporality of speech signals \cite{RNN_temporal}, but it requires high training costs \cite{RNN_cost}. \wxc{RNN cannot solve the problem of long time dependency?} 

Nevertheless, there are still some limitations in the proposed methods, including:
\begin{itemize}
\item[1)]
\textbf{The emotion causality in speech is not sufficiently explored.} 
The causality is a prerequisite for the perception of human~\cite{causal_perception}, which has a temporal priority in the time domain (cause precedes effect)~\cite{causal_time}. Besides, the speech signal always carries rich contextual sentimental information and the emotion with the same valence shares common antecedents and consequences~\cite{causal_emotion}. Therefore, the emotion causality is significant for addressing the SER task. Recently, various approaches based on emotion causality have been introduced to analyze emotions in diverse media. For instance, Mittal \textit{et al.}\cite{causal_relatedwork} analyzed the affective of movies with the multimodal method based on the ideas from emotion causation theories, which introduces Granger causality to model the temporal causality. These works illustrated the importance of causality in emotion analysis. However, to the best of our knowledge, most of works in the SER ignored the importance of emotion causality. 
\item[2)]
\textbf{The problem of long-term dependency persists.} 
%A major limitation of SER methods lies in that they do not include structures to capture high-level features. That is, they only contain the low-level convolution layers that are not effective to form the long-term dependencies. While the high-level convolution layer have larger receptive field, and only the higher-level layers are capable of building long-term dependencies. However, due to the small sample size problem in SER task, there is no effective solution to construct models with high-level convolution layers.
The existing SER methods did not well utilize high-level features. The low-level convolution layers are not as capable of building long-term dependencies as the high-level layers, which have a larger receptive field to maintain the long-term dependencies. Without the sufficient high-level features, most of the approaches failed to effectively build reliable long-term dependencies in the existing SER methods.

\item[3)]
\textbf{The existing single-scale architectures are inadequate for modelling speech emotion.} 
The human speech is not a type of single-scale signals. Instead, such signals contain ample information in nature, and should be treated as the multi-scale data. For instance, prosody has a multi-scale expression across the time domain, which results in abundant and subtle emotion changes in human speech~\cite{ms_emotion}. However, most of the existing approaches had been directed at modelling speech emotion on a global scale, suffering from the absence of local-scale modelling.
\end{itemize}

% 

%In addition, it is considered that the work on mining basic feature information is still far from well explored. The use of the single modality in SER to capture emotion representation is still preliminary. As the methods deploying a single feature for SER generally achieved low performance, many advanced works had used multi-modality methods to achieve higher classification accuracy~\cite{multimodel, multimodel1}. 

% However, to the best of our knowledge, there is no research in the SER field based on the TCNN.

%\wxc{Whether it needs to be deleted or not has little to do with our article}

To address these challenges, we propose a new approach based on TCNN, called Gated Multi-scale Temporal Convolutional Network (GM-TCNet). It aims to construct a novel emotional causality representation learning component with multi-scale receptive fields.

To the best of our knowledge, this is the first try to mine temporal causality among speech emotions in SER. The causality of different neurons is well maintained by introducing the causal convolution to the structure of GM-TCNet. Consequently, GM-TCNet can simulate the human perception of speech emotions by the causal convolution, and infer the emotions at the frame level. Specifically, we use the term ``emotion causality'' or ``temporal causality'' to refer to the constraint that speech should be processed in a forward manner. It is different from causal learning in~\cite{causal_learning}. 

Furthermore, by employing dilated convolution and gating mechanism, the ability of the dilated convolution is strengthened in building long-term sentimental dependency across the time domain. The gating mechanism can enhance the ability of low-level convolution layers to build a reliable long-term dependency. 

GM-TCNet utilizes skip connection to fuse high-level features with diverse receptive fields to capture abundant and subtle emotion changes in human speech. The proposed approach is evaluated on four commonly used datasets compared with the state-of-the-art (SOTA) approaches. The experimental results demonstrate the superior performance of the GM-TCNet, achieving great improvements on weighted average recall (WAR) and unweighted average recall (UAR) scores on four widely used datasets. The main contributions of this paper are summarized as follows.
%Our approach does not use other information (such as video, or text) or complex hand-crafted Low-Level Descriptors as the input, in which the design of salient features is highly dependent on the target corpus. We mainly focus on the extraction of high-level features. 

\begin{itemize}
\item[1)]
\textbf{A novel emotional causality representation learning component.} Constructed with dilated causal convolution layer and gating mechanism, the proposed GM-TCNet can capture the changes of emotion across time domain and better model the speech emotions at frame level. It also has a strong ability to build a reliable long-term sentimental dependency. To the best of our knowledge, this is the first attempt at applying the causality learning method to SER.
\item[2)]
\textbf{High-level features with multi-scale receptive fields.} GM-TCNet extracts high-level features from different Gated Convolution Blocks (GCB) with multi-scale receptive fields. It uses skip connection combining features to capture abundant and subtle emotion changes in human speech.
\item[3)]
\textbf{Superior-performance in speech emotion recognition.} The experimental results demonstrated that the proposed GM-TCNet can effectively produce the features with emotional causality from speech and significantly outperform the SOTA approaches.
\end{itemize}

The remainder of this paper is organized as follows. Section \ref{overview TCNN} gives a brief overview of TCNN, and the details of our approach are presented in Section \ref{proposed algorithm}. Section \ref{Experiment} reports experimental results along with discussions. Finally, Section \ref{Conclusions} concludes this paper and points out some future research directions.

\section{Overview of Temporal Convolutional Neural Networks}

\label{overview TCNN}

The original architecture of TCNN \cite{TCN1} is a sequential model to process data across the time domain. Formally speaking, the sequence model network is a function $F$ that produces a mapping: $X^{T+1}\rightarrow Y^{T+1}$. $F$ receives an input sequence $x_{0},x_{1},\ldots,x_{T}$ and produces a corresponding output sequence $y_{0},y_{1},\ldots,y_{T}$. As Eq.(\ref{E1}) shows, the target of training in $F$ is to evaluate $\hat{y_{0}},\hat{y_{1}},\ldots,\hat{y_{T}}$ by minimizing some loss functions between the corresponding output sequence $y_{0},y_{1},\ldots,y_{T}$ and the estimated sequence $\hat{y_{0}},\hat{y_{1}},\ldots,\hat{y_{T}}$. In SER, the input sequences are the low-level feature sequences, and the output sequences are the high-level feature sequences extracted from the time domain.
\begin{gather}
\hat{y_{0}},\hat{y_{1}},\ldots,\hat{y_{T}}=F(x_{0},x_{1},\ldots,x_{T})\label{E1}\\
O(s)=(x*_{d}f)(s)=\sum_{i=0}^{k-1}f(i)\cdot x_{s-d\cdot i}\label{E2}
\end{gather}

\begin{figure*}[t]
	\centering
	\includegraphics[scale= 0.35]{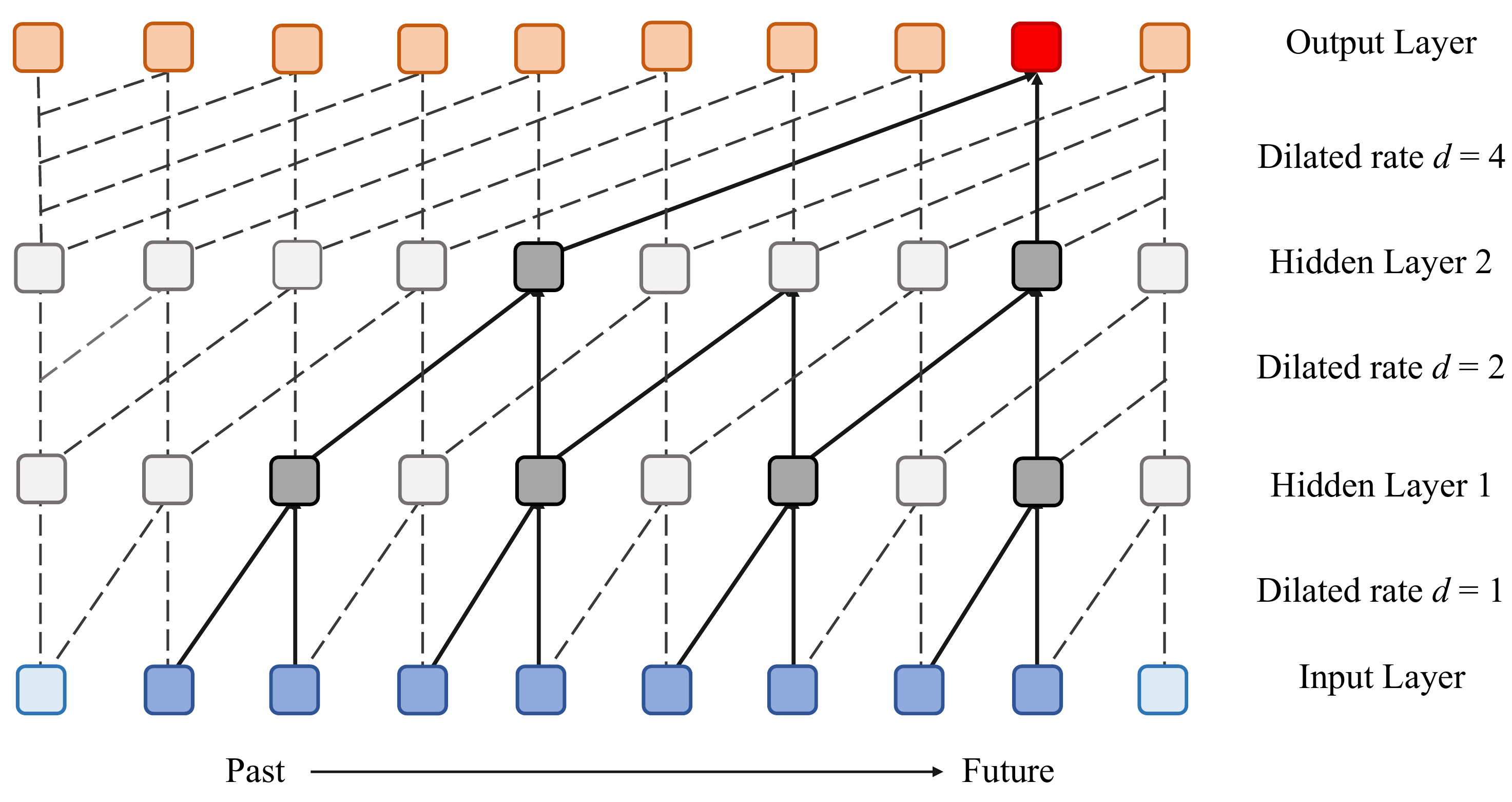}
	\caption{ A dilated causal convolution with dilated rates = 1,2,4 and kernel size k = 2 in the TCNN.}
	\label{Fig1}
\end{figure*}

TCNN deploys the dilated causal convolution layers to learn long-term dependencies and achieve causal constraints. The dilated convolution increases the receptive field exponentially. Formally, for a 1-D sequence input $x\in \mathbb{R}^{n}$ and a filter $f:\left \{0,\ldots,k-1 \right \}\rightarrow \mathbb{R}$, the dilated convolution operation $O$ on element $s$ of the sequence $x$ is defined by Eq.(\ref{E2}). $d$ is the dilated rate, $*_{d}$ represents the calculation symbol for dilated convolution, $k$ is the filter size, and ($s-d \cdot i$) accounts for the direction of the past. In Figure \ref{Fig1}, TCNN increases dilated rate $d$ exponentially with the depth of the network (i.e., $d = 2^{j}$ at level $j$ of the network). Moreover, the causal constraint indicates the prediction of $\hat{y_{t}}$ is only related to $x_{0},x_{1},\ldots,x_{t}$ and unrelated to the future inputs $x_{t+1},x_{t+2},\ldots,x_{T}$, which ensures that future information is not leaked to the past.

\begin{figure*}[t]
	\centering
	\includegraphics[width = 4cm]{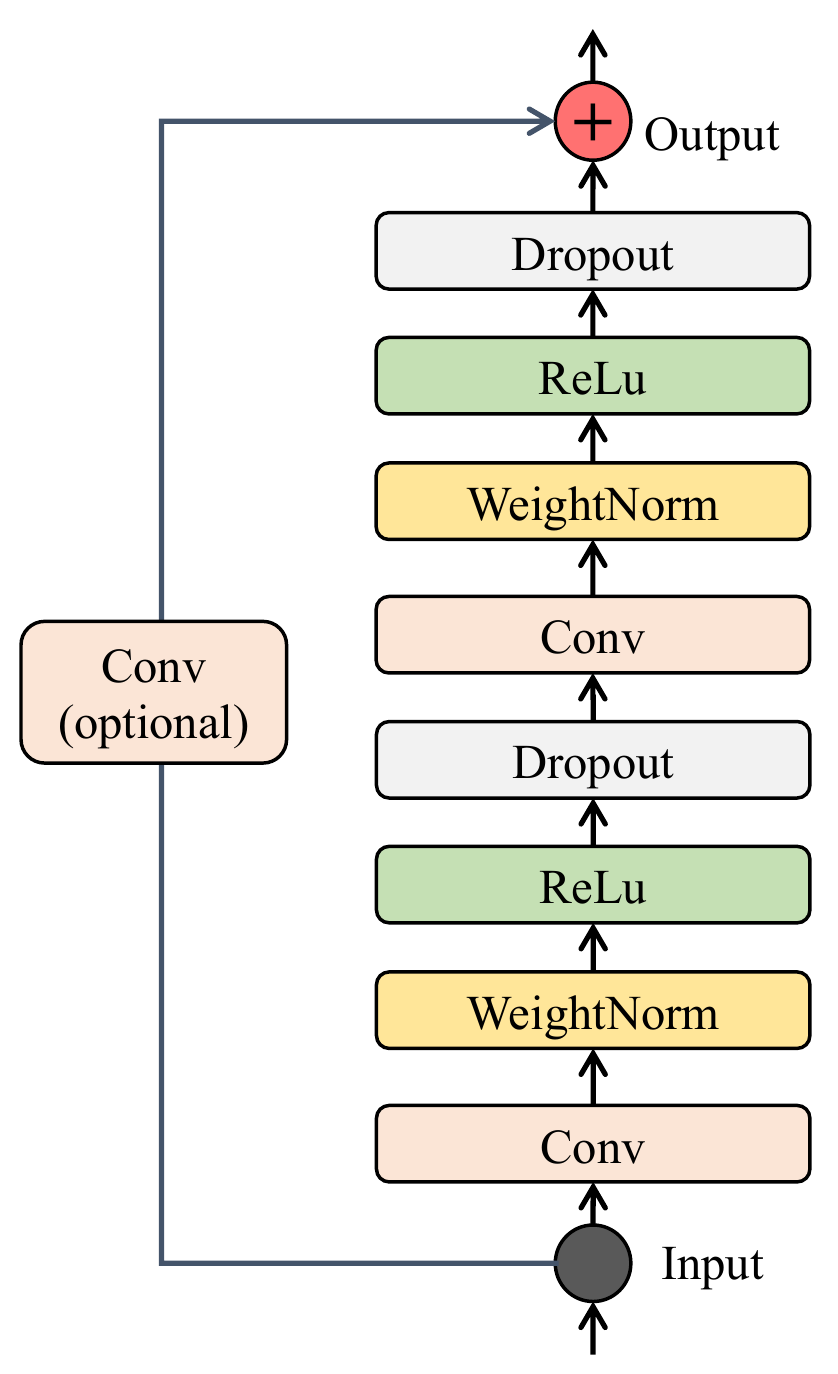}
	\caption{TCNN residual block. A $1 \times 1$ convolution layer is added when block input and output have a different shape.}
	\label{Fig2}
\end{figure*}

Another superiority of the TCNN is the usage of residual blocks \cite{resnet}, as shown in Figure \ref{Fig2}. The residual block is composed of two branches. The right one has an optional $1 \times 1$ convolution to ensure that the input and output have the same shape, and the left one is composed of two sets of identical blocks. Moreover, each block includes a dilated causal convolution layer, a weight regularization layer, a Rectified Linear Unit (ReLU) layer, and a spatial dropout layer.

Due to the excellent sequence modeling ability of TCNN, it has abundant applications in speech recognition. Pandey \textit{et al.} \cite{TCN_RW1} proposed an encoder-decoder architecture based on TCNN for real-time speech enhancement in the time domain. Zhang \textit{et al.} \cite{TCN_RW2} proposed another TCNN-based architecture with the gating mechanism for end-to-end monaural speech separation, which introduced the depth-wised convolution layers and dilated causal convolution layers but ignored the emotion causality in the speech. Furthermore, Tang \textit{et al.} \cite{TCN_RW3} applied dilated causal convolution layers instead of TCNN-based methods to tackle the SER task. They focused on the way to effectively learn the global and local representations through the max pooling layers and contextual stacking architecture, which ignored the importance of causal convolution in SER. 

Unlike the existing methods, to the best of our knowledge, our GM-TCNet is the first TCNN-based work for the SER task. The dilated causal convolutional layers are deployed to mine the causal relationship of different emotions, which are more oriented towards addressing long-term sentimental dependencies. In addition, the multi-scale architecture can capture abundant and subtle emotion changes in human speech and the novel gated residual block is deployed to change the dilated rate distribution. The details are given in the next section.

\section{The Proposed Approach}
\label{proposed algorithm}

This paper proposed a new TCNN-based approach called Gated Multi-scale Temporal Convolutional Neural Network (GM-TCNet). Compared to TCNN, GM-TCNet is designed by merging the gating mechanism and skip connection to control information flow and capture multi-scale temporal features. It first accepts the 39-D MFCC features as inputs to further extract high-level features through the Gated Temporal Convolutional Module (GTCM). And then the corresponding outputs are fed to the Global Average 1-D Pooling (GAP) and Fully Connected (FC) layer to produce the final decision.

\subsection{Feature Extraction}

\begin{figure*}[t]
	\centering
	\includegraphics[width=12.cm]{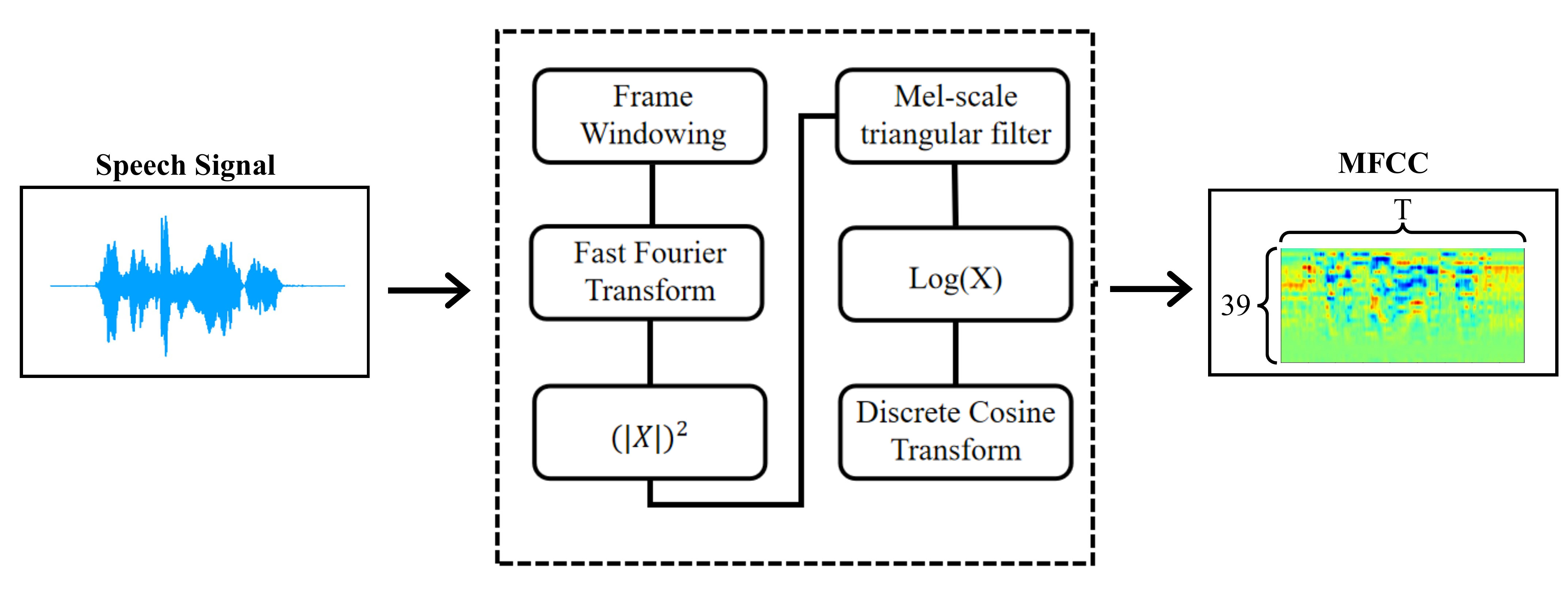}
	\caption{The workflow chart of MFCC feature extraction}
	\label{feature}
\end{figure*}

Figure \ref{feature} shows the MFCC feature extraction process. The framing and windowing operations are first applied to each speech data. Then, each frame signal performs a fast Fourier transform to obtain the spectrum. Next, the related modulus and their square are calculated based on the speech signal spectrum to generate the power spectrum, which passes through a set of Mel-scale triangular filter banks. Finally, the logarithmic energy output is processed by the discrete cosine transformation to obtain the MFCC features. These coefficients are spliced together and transformed to a set of 39-D features, which serve as the inputs to GM-TCNet.

\subsection{GM-TCNet}

\begin{figure*}[t]
	\centering
	\includegraphics[width = 14cm]{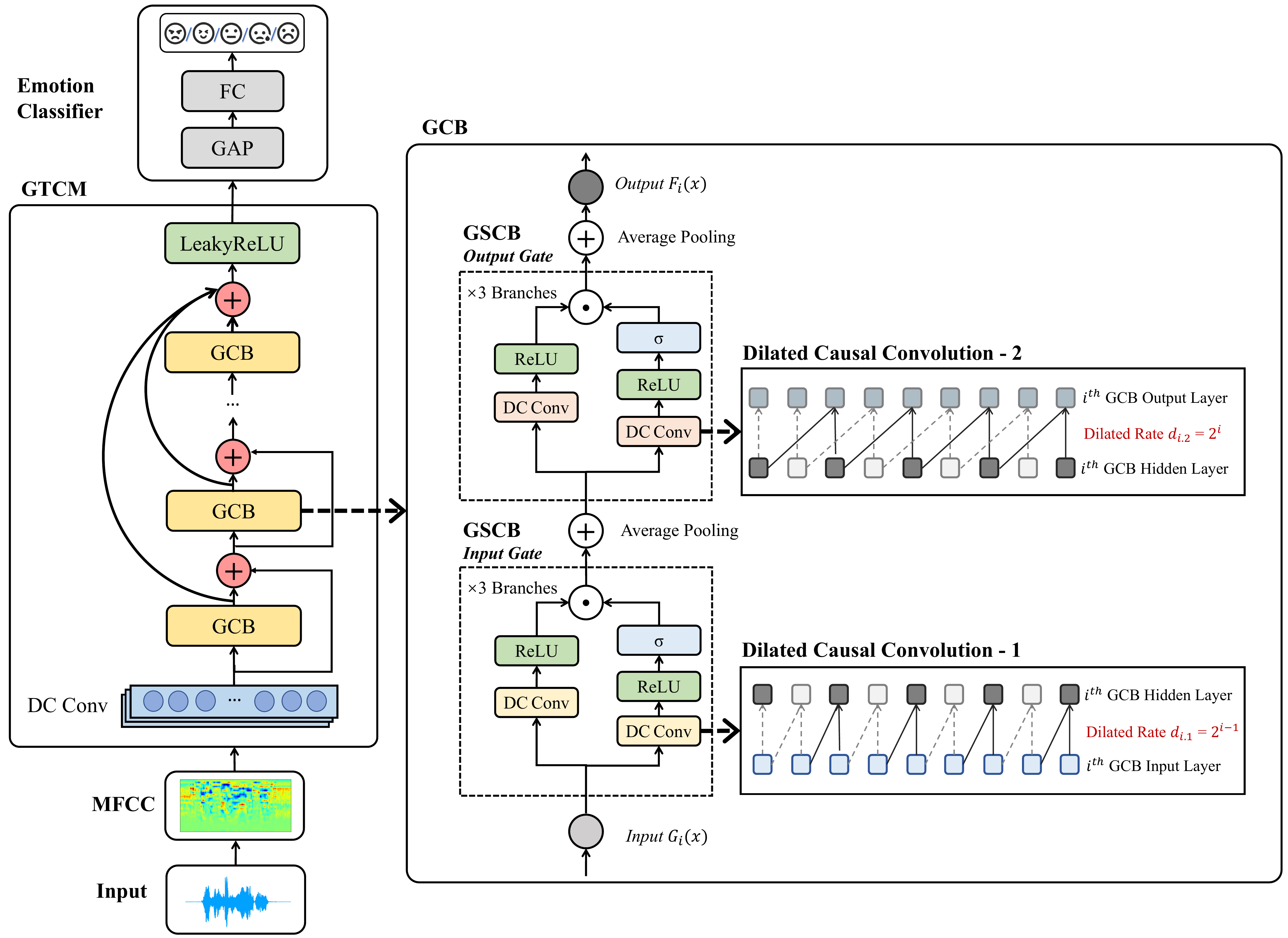}
	\caption{The structure of the GM-TCNet is composed of the GTCM module (including a Dilated Causal Convolutional layer and seven GCBs), and the emotion classifier (including the Global Average 1-D Pooling and Fully Connected layer).}
	\label{Strucre}
\end{figure*}

%Our network, named Gated Multi-scale Temporal Convolutional Network (GM-TCNet for short), is proposed based on the TCNN. It is composed of two parts, as shown in Figure \ref{Strucre}: (1) the GTCM, extracting the high-level features from the MFCC features; (2) the emotion classifier, generating the final classification decision. 

%The network architecture is mentioned in 3.3.1, the gated convolutional block is mentioned in 3.3.2, the gated sub-residual block is mentioned in 3.3.3, the output layer is mentioned in 3.3.4.

\subsubsection{Gated Temporal Convolutional Module}

Gated Temporal Convolutional Module(GTCM) is the core of GM-TCNet. It is built by stacking a 1-D causal convolutional layer and seven Gated Convolutional Blocks (GCB) with the exponentially increasing dilated rates. There are two levels in a GCB, each of which has three branches with the same architecture called Gated Sub Convolution Block (GSCB). As shown in Figure \ref{Strucre}, for the $i^{th}$ GCB, the dilated rate $d_{i.1}$ in the first level is $2^{i-1}$ and the dilated rate $d_{i.2}$ of the three sub-blocks ensembled in the second level is $2^i$. For example, the dilated rates of the first level and the second level GSCBs in the first GCB are 1 and 2 respectively, and those of the first level and the second level GSCBs in the seventh GCB are 64 and 128 respectively.

GTCM accepts a 2-D feature with the size of $T \times 39$ as inputs, where $T$ represents the number of frames and 39 represents the dimension of the MFCC features. The first layer of GTCM performs 1-D causal convolution with the kernel size of 1 and filter number of 39. Then the output of the first layer is fed to seven GCBs sequentially. The skip connection sums the outputs $F_i(x)$ from seven GCBs $(i=1,2,\cdots,7)$ and the summed output is fed to the LeakyReLU layer (alpha=0.05). In particular, the first six GCBs use residual connections to effectively learn modifications to the identity mapping~\cite{resnet}. The output of GTCM is fed to the emotion classifier at last.

\subsubsection{Gated Convolution Block}
Based on the residual block in the TCNN structure, each GCB in GTCM introduces the gating mechanism and average pooling strategy. The gating mechanism provides more effective control on information transmission and the capability of learning long-term sentimental dependencies~\cite{gatenet}. The average pooling strategy aids in improving each single model's performance. Specifically, as Eq.(\ref{E3}) shows, the input $G_{i}(x)$ of the $i^{th}$ GCB is equal to $H_{i-1}(x)$ when $i>1$. When $i=1$, the input $G_{1}(x)$ of the first GCB comes from the output of the first 1-D convolutional layer, represented by $x$. Moreover, as Eq.(\ref{E6}) shows, $H_{i}(x)$ is the sum of the $G_{i}(x)$ and the output $F_{i}(x)$ of the $i^{th}$ GCB. 

Figure \ref{Strucre} shows that GCB is divided into two levels. The first level is called "Input Gate", used to identify the importance of the current inputs to the second level. It also helps to determine how much the input of the current GCB is preserved to the unit state $C_{i}$. The second level controls how much $C_{i}$ is captured in the output $F_{i}(x)$ of the $i^{th}$ GCB, named "Output Gate".
\begin{gather}
G_{i}(x)=\begin{cases}
x &  i=1 \\ 
H_{i-1}(x) &  2\leq i \leq 7
\end{cases}\label{E3}\\
I_i^j(x)={\rm ReLU}(\mathbf{W}_{2}*_d G_{i}(x)) \odot \sigma({\rm ReLU}(\mathbf{W}_{2}*_d G_{i}(x))), 1\leq j\leq 3\label{E_I}\\
C_{i}(x)=\frac{1}{3}(I_i^{1}(x)+I_i^{2}(x)+I_i^{3}(x))\label{E4}\\
O_i^j(x)={\rm ReLU}(\mathbf{W}_{2}*_d C_{i}(x) ) \odot \sigma({\rm ReLU}(\mathbf{W}_{2}*_d C_{i}(x) )), 1\leq j\leq 3\label{E_O}\\
F_{i}(x)=\frac{1}{3}(O_i^{1}(x)+O_i^{2}(x)+O_i^{3}(x))\label{E5}\\
H_{i}(x)=F_{i}(x)+G_{i}(x)\label{E6}
\end{gather}

Each level has three GSCBs, which share the same structure. As shown in Figure \ref{Strucre}, the input feature is fed to two different branches respectively, which first performs a 1-D dilated causal convolution with a dilated rate $d$ and goes through the ReLU layer. The right branch still passes through the Sigmoid layer. The output range of the right branch takes values in the range of (0,1) because of the Sigmoid layer. As the final output is multiplied element-wise by the production of two branches, it can provide the overall importance of the current input. Specifically, as Eq.(\ref{E_I}) and Eq.(\ref{E_O}) show, $\mathbf{W}_{2}$ denotes the weight of the dilated causal convolution and the kernel size is 2, $*_d$ is the dilated causal convolution operation with the dilated rate $d$, $\sigma(\cdot)$ is the sigmoid function, $\rm{ReLU}(\cdot)$ is the ReLU activation function, $I_i^j(x)$ and $O_i^j(x)$ denote $j^{th}$ GSCB of input gate and output gate in the $i^{th}$ GCB. 

Furthermore, each GCB employs the average pooling strategy to improve the model's performance. In the $i^{th}$ GCB, the dilated rate is $2^{i-1}$ for the input gate, and their outputs $I_i^{1}(x), I_i^{2}(x), I_i^{3}(x)$ are averaged as $C_i(x)$ by Eq.(\ref{E4}). $C_i(x)$ is sent to the output gate, whose dilated rate $d$ is $2^{i}$. After that, the same steps are processed by Eq.(\ref{E_O}) to get the output results $O_i^{1}(x), O_i^{2}(x),O_i^{3}(x)$ and then the output $F_{i}(x)$ is averaged by Eq.(\ref{E5}). To effectively learn modifications to the identity mapping rather than the entire transformation, the residual connection is introduced in Eq.(\ref{E6}). $H_{i}(x)$ is the sum of the input $G_{i}(x)$ and the output $F_{i}(x)$ of the $i^{th}$ GCB.

\subsubsection{Emotion Classifier}

\begin{figure*}[t]
	\centering
	\includegraphics[width = 8.0cm]{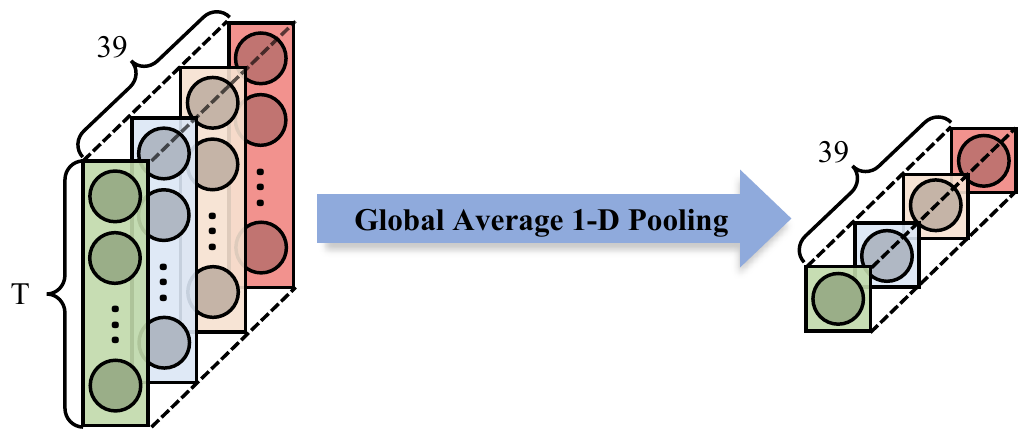}
	\caption{The process of data transfer in Global Average 1-D Pooling Layer.}
	\label{Fig_GAP}
\end{figure*}

The process of the emotion classifier is shown in Figure \ref{Strucre}, which includes the GAP layer and FC layer. After fusing the output of GTCM, the network uses the GAP layer to reduce the number of learnable parameters and drives CNN to fit the inputs at modified size \cite{GAP1}. As depicted in Figure \ref{Fig_GAP}, GAP generates a feature map for each related object with the high-level features generated by seven GCBs. Next, it takes the average of each feature map in the time domain to avoid overfitting. Finally, the generated vector is sent to the FC layer.

The FC layer uses the Softmax function as the activation function, which handles the high-level features extracted by GM-TCNet for the classification task. It maps the outputs of multiple neurons to [0,1], and normalizes them in Eq.(\ref{E_softmax}). Among them, $e^i$ represents the exponential function of the $i^{th}$ element of the high-level features input by GM-TCNet, and $s(z)_i$ denotes the softmax value.
 \begin{gather}
s(z)_{i}=\frac{e^{z_i}}{\sum_{j=1}^{K} e^{z_j}} \text{ for } i=1,...,K \text{ and } z=(z_1,z_2,...,z_K)\in \mathbb{R}^K\label{E_softmax}
\end{gather}

\section{Experiment}

\label{Experiment}
\subsection{Experiment Settings}
\paragraph{Dataset}

Four public emotion datasets are used in the experiments: the Institute of Automation of Chinese Academy of Sciences (CASIA) \cite{database1}, Berlin Emotional dataset (EMODB) \cite{database2}, Ryerson Audio-Visual dataset of Emotional Speech and Song (RAVDESS) \cite{database3}, and Surrey Audio-Visual Expressed Emotion dataset (SAVEE) \cite{database4}. The language used in both RAVDESS and SAVEE is English, while the speeches in EMODB and CASIA datasets are in German and Chinese respectively. The details of these datasets are given in Table \ref{tab_2.1} and Table \ref{tab_2.2}.
\begin{table}[h]
\small
\renewcommand{\arraystretch}{1.5}
\centering
% \begin{tabular}{cccc>{\centering}p{5cm}c}
\begin{tabular*}{1\linewidth}{@{\extracolsep{\fill}}rrcclc}
\toprule[1.5pt]
Dataset & Language & Speakers & Numbers & Emotion & Frequency\\
\midrule
CASIA & Chinese & \tabincell{l}{2 males, 2 females} & 1200 & \tabincell{l}{Angry, Fear, Happy,\\ Neutral, Sad, Surprise} & 22.1 kHz\\
EMODB & German & \tabincell{l}{5 males, 5 females}& 535 & \tabincell{l}{Angry, Boredom, Disgust, Fear,\\ Happy,  Neutral, Sad} & 16.0 kHz\\
RAVDESS & English & \tabincell{l}{12 males, 12 females} & 1440 & \tabincell{l}{Angry, Calm, Disgust, Fear,\\ Happy, Neutral, Sad, Surprise} & 48.0 kHz\\
SAVEE & English & \tabincell{l}{4 males} & 480 & \tabincell{l}{Angry, Disgust, Fear, Happy, \\ Neutral, Sad, Surprise} & 44.1 kHz\\
\bottomrule[1.5pt]
\end{tabular*}
\caption{The detailed information of speech emotion datasets.}
\label{tab_2.1}
\end{table}

% \begin{table}[t]
% \footnotesize
% \renewcommand{\arraystretch}{1.5}
% % \setlength{\tabcolsep}{1.65mm}
% \centering
% \begin{tabular}{cccccccccc}
% \toprule
% Dataset & Neutral & Anger & Disgust & Fear & Happy & Sad & Surprise & Boredom & Calm\\
% \midrule
% RAVDESS & 96 & 192 & 192 & 192 & 192 & 192 & 192 & – & 192\\
% EMODB & 79 & 127 & 46 & 69 & 71 & 62 & – & 81 & –\\
% SAVEE & 120 & 60 & 60 & 60 & 60 & 60 & 60 & – & –\\
% CASIA & 200 & 200 & – & 200 & 200 & 200 & 200 & – & –\\
% \bottomrule
% \end{tabular}
% \caption{The details of data distributions in four datasets }
% \label{tab_2.2}
% \end{table}

\begin{table*}[htbp]
\centering
\small
\renewcommand{\arraystretch}{1.5}
\begin{tabular*}{1\linewidth}{@{\extracolsep{\fill}}r*9{c}}
\toprule[1.5pt]
Dataset & Angry & Boredom & Calm & Disgust & Fear & Happy & Neutral & Sad & Surprise\\
\midrule
CASIA & 200 & - & - & - & 200 & 200 & 200 & 200 & 200\\

EMODB & 127 & 81 & - & 46 & 69 & 71 & 79 & 62 & -\\

RAVDESS & 192 & - & 192 & 192 & 192 & 192 & 96 & 192 & 192\\

SAVEE & 60 & - & - & 60 & 60 & 60 & 120 & 60 & 60\\
\bottomrule[1.5pt]
\end{tabular*}
\caption{The details of data distributions in four datasets.}\label{tab_2.2}
\end{table*}

\paragraph{Features}
In the experiments, the 39-D MFCC features are extracted by the Librosa toolbox \cite{librosa} with the default settings. That is, the frame length is 0.05 s, the frame shift is 0.0125 s, the sample rate is 22050 Hz and the window function added for the speech data is Hamming window.

\paragraph{Implementation and Training}

The proposed approach is implemented in the TensorFlow framework \cite{tensorflow}. The batch size is set to 64 and the training process is optimized by Adam algorithm \cite{adam} with an initial learning rate $\alpha$ = $1.0\times 10^{-3}$, exponential decay rates $\beta_1$ = 0.93, $\beta_2$ = 0.98, and weight decay $\epsilon$ = $1.0\times 10^{-8}$. Moreover, the cross-entropy loss is employed as the loss function. For better comparison with the SOTA approaches, the hold-out validation (80\% data for training and 20\% for testing), 5-fold cross-validation (CV), and 10-fold CV schemes are all used. To compare with other methods fairly, we used all data in all comparison experiments with random divisions. In each type of partitioning, we performed multiple partitions and verified the experimental results.

\paragraph{Evaluation Metrics}

The Weighted Average Recall (WAR) is the weighted average recall with weights equal to the class probabilities, and Unweighted Average Recall (UAR) is the average recall of different sentiment classes. They are employed for performance comparison, as defined by:

\begin{equation}
\begin{aligned}
W A R=\sum_{\alpha=1}^{K} \frac{M}{N}\times \frac{\sum_{\beta=1}^{M} T P_{\alpha}^{\beta}}{\sum_{\beta=1}^{M} (T P_{\alpha}^{\beta}+ F N_{\alpha}^{\beta})}
\end{aligned}
\label{WAR}
\end{equation}
\begin{equation}
%\hspace{1em}
U A R=\frac{1}{K} \sum_{\alpha=1}^{K} \frac{\sum_{\beta=1}^{M} T P_{\alpha}^{\beta}}{\sum_{\beta=1}^{M} (T P_{\alpha}^{\beta}+F N_{\alpha}^{\beta})}
\label{UAR}
\end{equation}

Here, $K$, $M$ and $N$ represent the number of sentiment classes, the number of speech signals of class $\alpha$ and the number of all speech signals respectively. $T P_{\alpha}^{\beta}, T N_{\alpha}^{\beta}, F P_{\alpha}^{\beta} \text { and } F N_{\alpha}^{\beta}$ represent the true positive, true negative, false positive, and false negative values of class $\alpha$ for speech signal $\beta$ respectively.

\subsection{Experimental Results}
\begin{table}
\small
\centering
\renewcommand{\arraystretch}{1.5}
% \begin{tabular}{cccccc}
\begin{tabular*}{1\linewidth}{@{\extracolsep{\fill}}lccccc}
\toprule[1.5pt]
\textbf{Split ratio }  & \textbf{Metrics} & \textbf{CASIA} & \textbf{EMODB} & \textbf{RAVDESS} & \textbf{SAVEE} \\ \midrule
\multirow{2}{*}{8:2   hold-out}         & WAR       & 92.50 & 95.33 & 90.28   & 90.63 \\
                                        & UAR       & 92.21 & 95.66 & 90.03   & 91.04 \\\midrule
\multirow{2}{*}{5-fold   CV (Max)}      & WAR       & 89.50 & 89.35 & 87.08   & 84.79 \\
                                        & UAR       & 89.50 & 89.47 & 86.91   & 83.33 \\
\multirow{2}{*}{5-fold   CV (Average)}  & WAR       & 88.68 $\pm$ 0.75 & 88.97 $\pm$ 0.30 & 86.92 $\pm$ 0.18  & 83.63 $\pm$ 0.73\\
                                        & UAR       & 88.68 $\pm$ 0.75 & 88.83 $\pm$ 0.49 & 86.71 $\pm$ 0.35  & 81.90 $\pm$ 0.88\\\midrule
\multirow{2}{*}{10-fold   CV (Max)}     & WAR       & 90.17 & 91.40 & 87.64   & 86.01 \\
                                        & UAR       & 90.17 & 90.45 & 87.30   & 84.40 \\
\multirow{2}{*}{10-fold   CV (Average)} & WAR       & 89.35 $\pm$ 0.59 & 91.06 $\pm$ 0.28 & 86.83 $\pm$ 0.55  & 85.10 $\pm$ 0.52\\
                                        & UAR       & 89.35 $\pm$ 0.59 & 90.28 $\pm$ 0.28 & 86.56 $\pm$  0.54 & 83.47 $\pm$ 0.55\\ 
\bottomrule[1.5pt]
\end{tabular*}
\caption{Performance (\%) of the proposed GM-TCNet on various datasets. }
\label{TableAC}
\end{table}

In the experiments, four speech emotion datasets CASIA, EMODB, RAVDESS, and SAVEE are used to verify the effectiveness of the GM-TCNet. Table \ref{TableAC} shows the results of our proposed model. Moreover, we list the highest accuracy obtained by different studies in recent publications in Tables \ref{tab_casia}-\ref{tab_savee}. We can see from these tables that GM-TCNet performs better than other approaches in most cases and obtains the highest accuracy on all four datasets. These results confirm the high performance of GM-TCNet across different datasets.

\begin{figure}[!h]
    \centering
    \subfloat[CASIA]{
    \begin{minipage}[t]{0.45\linewidth}
    \centering
    \includegraphics[width = 7.5cm]{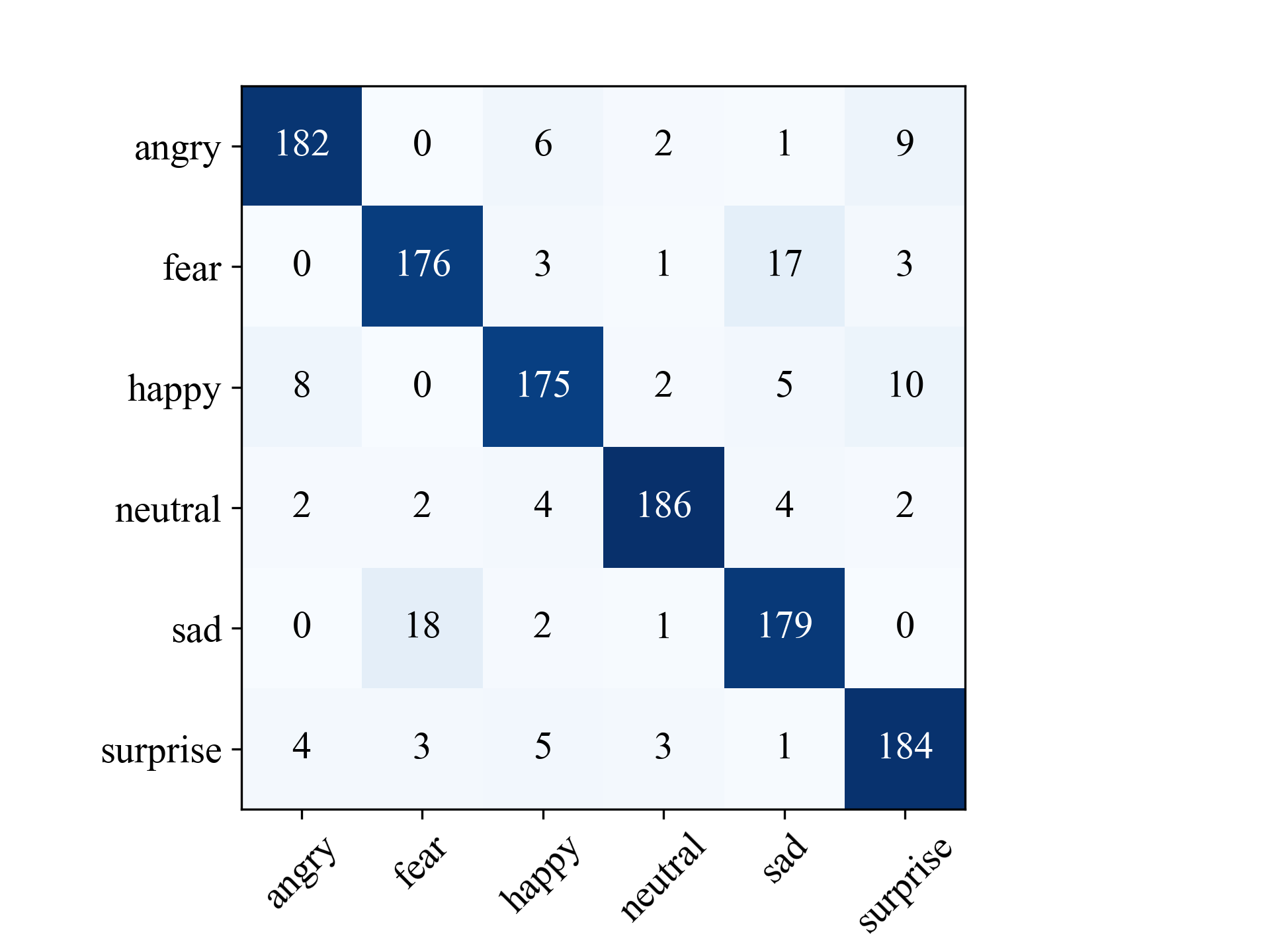}
    % \caption{fig2}
     \vspace{-40mm}
    \end{minipage}%
    \label{CASIA_matrix}
    }
    \subfloat[EMODB]{
    \begin{minipage}[t]{0.45\linewidth}
    \centering
    \includegraphics[width = 7.5cm]{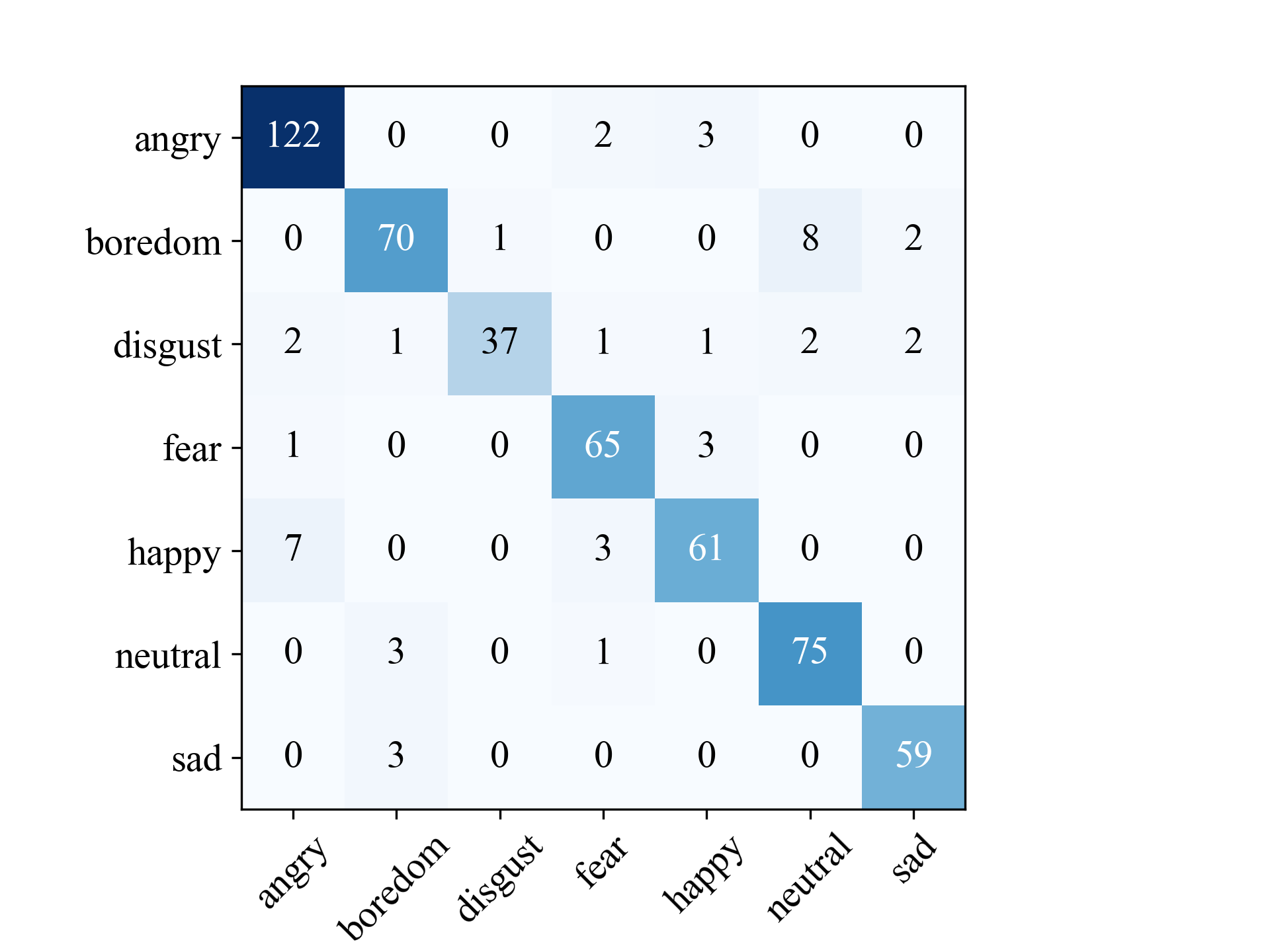}
    \vspace{-40mm}
    \end{minipage}
    \label{EMODB_matrix}
    }
    
    \subfloat[RAVDESS]{
    \begin{minipage}[t]{0.45\linewidth}
    \centering
    \includegraphics[width = 7.5cm]{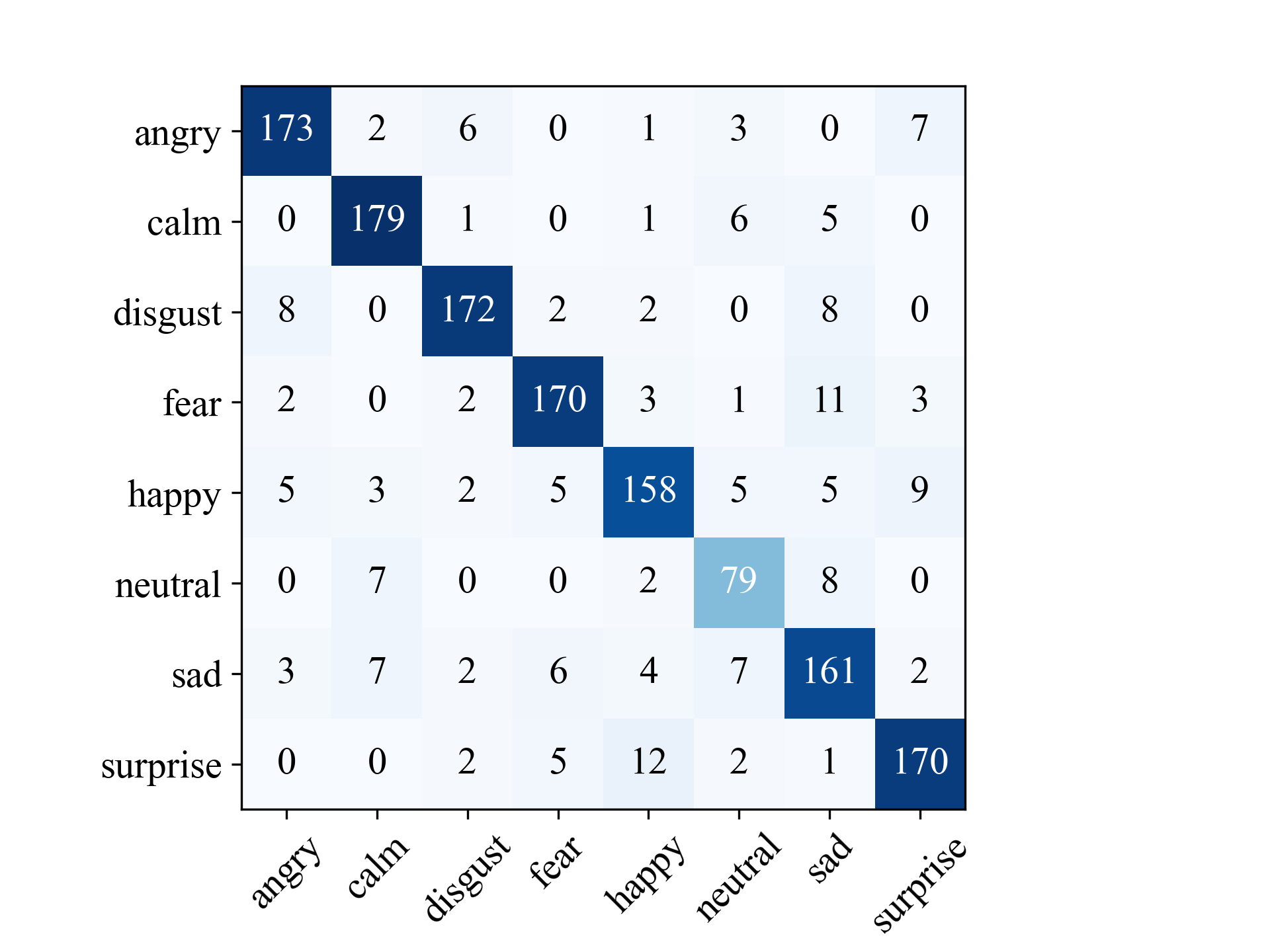}
    %\caption{fig2}
     \vspace{-40mm}
    % \label{Fig4.6}
    \end{minipage}
    \label{RAVDE_matrix}
    }
    \subfloat[SAVEE]{
    \begin{minipage}[t]{0.45\linewidth}
    \centering
    \includegraphics[width = 7.5cm]{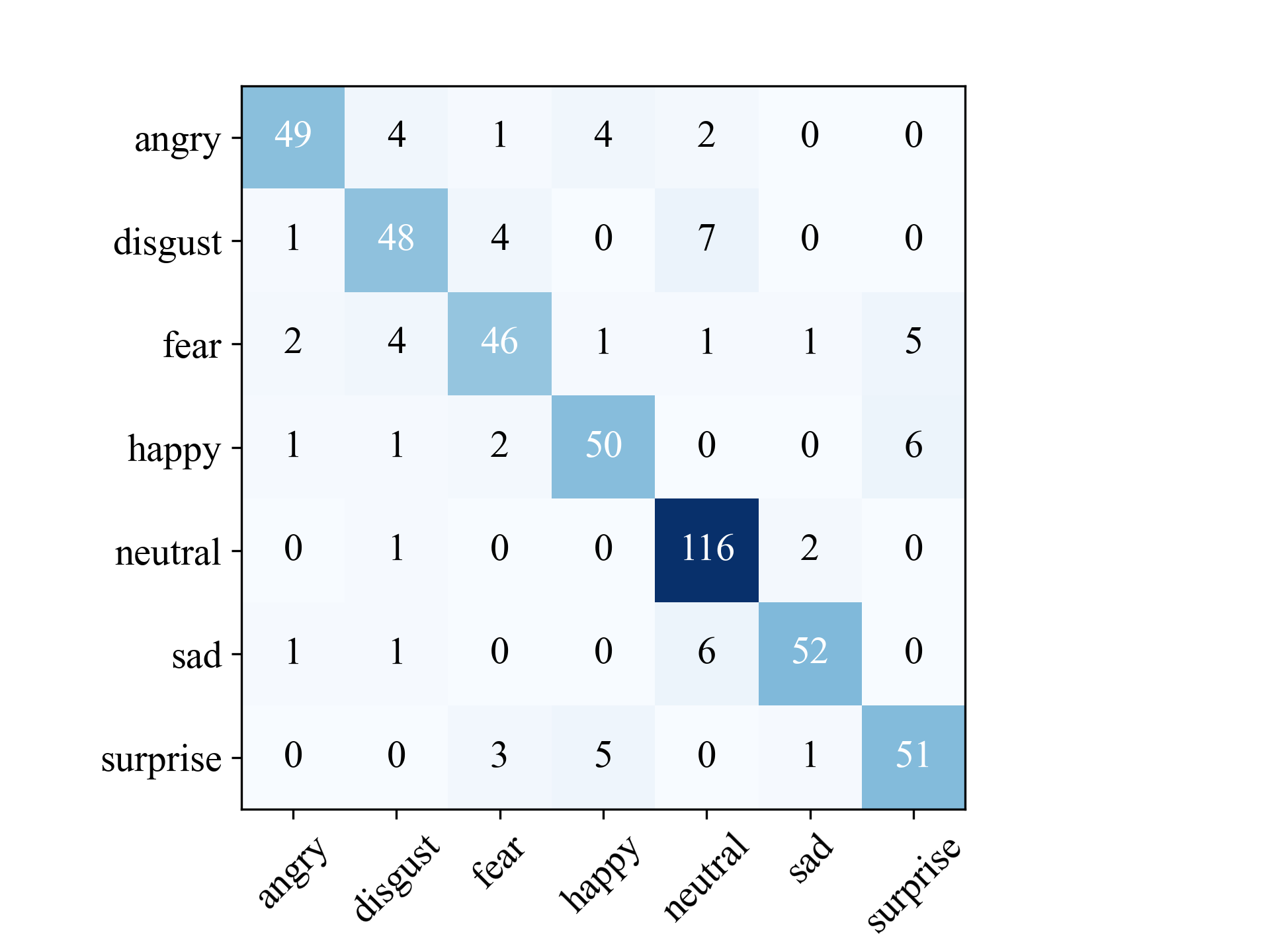}
    %\caption{fig2}
     \vspace{-40mm}
    % \label{Fig4.8}
    \end{minipage}
    \label{SAVEE_matrix}
    }
\centering
\caption{The 10-fold CV confusion matrix obtained using GM-TCNet on various datasets.}
\label{Matrix}
\end{figure} 

\subsubsection{CASIA}

GM-TCNet obtains an 89.50\% WAR score on the CASIA dataset, which is 1.60\% higher than the highest accuracy ever reported in 5-fold CV. Hong \textit{et al.} \cite{sota8} only used the MFCC feature for SER and achieved 83.65\% WAR score. With the same type of MFCC feature, our model gains a 8.85\% higher WAR score on the CASIA dataset. Overall, GM-TCNet improves the performance to 92.50\%, 89.50\%, 90.17\% on the hold-out, 5-fold CV and 10-fold CV, respectively. Furthermore, as shown in Figure \ref{CASIA_matrix}, GM-TCNet achieves 93.00\% and 91.00\% accuracy on the Neutral and Angry emotions, respectively, getting the best scores ever reported.

\begin{table*}[!h]
% \footnotesize
\small
\renewcommand{\arraystretch}{1.5}
\centering
\begin{tabular*}{1\linewidth}{@{\extracolsep{\fill}}lcllcc}
\toprule[1.5pt]
Study            & Year & Method                                          & Split ratio              & WAR & UAR \\
\midrule
L. Sun \textit{et al.} \cite{sota3}       & 2019 & Decision Tree SVM     & 10-fold CV & 85.08       & 85.08                 \\
M. Gao \textit{et al.} \cite{sota4}       & 2019 & CNN                                             & 5-fold CV  & 87.90        & 87.90                 \\
Z. Hong \textit{et al.} \cite{sota8}      & 2020 & LCNN                                            & 8:1:1 hold-out           & 83.65       & -                 \\
L. Chen \textit{et al.} \cite{sota7}      & 2021 & Two-layer Fuzzy Multiple Random Forest  & 5-fold CV  & 85.83       & 85.83                 \\
J. He \textit{et al.} \cite{sota32} & 2021 & CNN and BLSTM       & 7:3 hold-out            & 74.17       & -                 \\
\midrule
Our proposed & 2022 & GM-TCNet & 8:2 hold-out & \textbf{92.50}	& \textbf{92.21} \\
Our proposed & 2022 & GM-TCNet & 5-fold CV & \textbf{89.50}	& \textbf{89.50} \\
Our proposed & 2022 & GM-TCNet & 10-fold CV & \textbf{90.17}	& \textbf{90.17} \\
\bottomrule[1.5pt]
\end{tabular*}
% }
\caption{The performances of different approaches on the CASIA dataset.}
\label{tab_casia}
\end{table*}

\subsubsection{EMODB}

%On the EMODB dataset, GM-TCNet achieves a 96.26\% accuracy, and its confusion matrix is shown in Figure \ref{Fig4.4}. In detail, our network has achieved 100\% accuracy on the Angry, Boredom and Disgust emotions, and achieved higher than 88.89\% accuracies in all emotions. Ilyas OZER\cite{sota19} used CNN's based method to obtain the accuracy of 91.32\%, which was the highest accuracy result on the EMODB to the best of our knowledge. Our method outperforms theirs and gets a 4.94\% higher accuracy. At the same time, our accuracy and UAR scores are higher than the methods proposed by Tuncer \textit{et al.}\cite{sota17} by 6.17\% and 6.28\%. It shows that our algorithm can achieve the best overall performance and better balanced results.

%Compare with other methods, Table \ref{tab_emodb} shows that GM-TCNet achieves the highest accuracy of 96.26\% and the results on the Angry, Disgust, Fear and Neutral emotions are the highest. Therefore, it is concluded that our method leads to well balanced performance.

On the EMODB dataset, Table \ref{tab_emodb} shows that GM-TCNet achieves 89.35\% and 91.40\% WAR scores in the 5-fold and 10-fold CV. As Figure~\ref{EMODB_matrix} shows that our model achieved 96.06\% and 95.16 \% accuracy on the Angry and Neutral emotions, higher than the results of other emotions. Ozer~\cite{sota19} used CNN based method to obtain 91.32\% accuracy index on the 10-fold CV, which was the highest accuracy result reported on the EMODB dataset. Nevertheless, our approach is slightly higher than theirs in terms of performance. At the same time, our accuracy and UAR scores are 1.31\% and 0.98\% higher than the methods proposed by Tuncer \textit{et al.}~\cite{sota17}. It shows that our approach can achieve the best overall performance and better balanced results. Compared with the approach of generic recurrent architectures (BiLSTM, etc.), our approach can trace long-term dependencies to some extent. For instance, our approach obtains a 3.90\% improvement on UAR scores compared with BiLSTM approach~\cite{sota12}. It indicates that our method can build a long-term dependency across the time domain, which maintains the temporal information of the speech as well.

\begin{table*}[!h]
\centering
\small
\renewcommand\arraystretch{1.5}
\begin{tabular*}{1\linewidth}{@{\extracolsep{\fill}}lcllcc}
\toprule[1.5pt]
Study            & Year & Method                                          & Split ratio              & WAR & UAR \\
\midrule
T. Özseven \cite{sota26}                                 & 2019 & SVM                                             & 10-fold CV          & 84.62       & -                 \\
G. Assunção \textit{et al.} \cite{sota20}                               & 2020 & Logistic Model Tree                             & 5-fold CV           & 80.40 & 80.00             \\
F. Daneshfar and S. J. Kabudian \cite{sota11} & 2020 & DNN                                             & Leave-OneSpeaker-Out     & 82.82       & -                 \\
Mustaqeem \textit{et al.} \cite{sota12}                               & 2020 & BiLSTM                                          & 5-fold CV           &    -    & 85.57                \\
D. Issa \textit{et al.} \cite{sota13}                               & 2020 & CNN                                             & 5-fold CV           & 86.10        & -                 \\
C. A. Jason \textit{et al.} \cite{sota24}             & 2020 & ANN                                             &  - & 86.20        & -                 \\
L. Kerkeni \textit{et al.} \cite{sota14}                           & 2019 & SVM                                             & 10-fold CV          & 86.22       & -                 \\
S. Yildirim \textit{et al.} \cite{sota15}                         & 2021 & SVM                                             & 10-fold CV          &    -   & 78.89                \\
L. Chen \textit{et al.} \cite{sota7}                           & 2020 & TLFMRF & 5-fold CV           & 87.85       & -                 \\
T. Tuncer \textit{et al.} \cite{sota17}                           & 2021 & SVM                                             & 10-fold CV          & 90.09 & 89.47            \\
W. Zehra \textit{et al.} \cite{sota33}                            & 2021 & SMO           & Leave-OneSpeaker-Out     & 90.40        & -                 \\
I. Ozer \cite{sota19}                                     & 2021 & CNN                                             & 10-fold CV          & 91.32       & -                 \\
\midrule
Our proposed & 2022 & GM-TCNet & 8:2 hold-out & \textbf{95.33}	& \textbf{95.66} \\
Our proposed & 2022 & GM-TCNet & 5-fold CV & \textbf{89.35}	& \textbf{89.47} \\
Our proposed & 2022 & GM-TCNet & 10-fold CV & \textbf{91.40}	& \textbf{90.45} \\
\bottomrule[1.5pt]
\end{tabular*}
% }
\caption{The performances of different approaches on the EMODB dataset.}
\label{tab_emodb}
\end{table*}

\subsubsection{RAVDESS}

%On the RAVDESS dataset, GM-TCNet obtains the highest accuracy with 2.85\% higher than the second best method. As shown in Figure \ref{Fig4.6}, the accuracies of the Calm, Disgust and Fear emotions are higher than 95\%. At the same time, accuracies and UAR scores are 2.85\% and 2.60\% higher than the method proposed by Tuncer \textit{et al.}\cite{sota17}. Compared with the methods using the MFCC feature (\cite{sota21} and \cite{sota24}), our method promotes the results by 12.08\% and 10.07\% on the accuracy respectively. It shows that our method makes full use of the temporal information in the MFCC feature, which produces a positive effect on the SER. Similarly, comparing the results obtained by MFCC feature and 1-D CNN\cite{sota31}, ours gets 15.03\% and 15.03\% higher accuracy and UAR scores.

%In short, Table \ref{tab_ravdess} shows that GM-TCNet achieves the highest accuracy and UAR results, and also obtains the highest accuracy on the Calm, Disgust, Fear, and Happy emotions.

Table \ref{tab_ravdess} shows that GM-TCNet achieves the highest WAR and UAR scores on the RAVDESS dataset. In detail, GM-TCNet obtains the highest WAR scores in all three data split schemes. As shown in Figure~\ref{RAVDE_matrix}, the accuracies of the Angry and Calm emotions are higher than 90\%. At the same time, the WAR and UAR scores of our proposed method are close to those of the method proposed by Tuncer \textit{et al.}~\cite{sota17}. Compared with the method using the MFCC feature~\cite{sota13}, our method promotes the results by 15.47\% on the WAR scores. It shows that our method is able to make full use of the temporal information in the MFCC feature, which positively improves its performance of the SER.

\begin{table*}[!h]
\centering
\small
\renewcommand\arraystretch{1.5}
\setlength{\tabcolsep}{0.5mm}{
% \begin{tabular}{p{3.3cm}p{0.8cm}p{5.1cm}p{3.3cm}p{1.3cm}p{1.2cm}}
\begin{tabular*}{1\linewidth}{@{\extracolsep{\fill}}lcllcc}
\toprule[1.5pt]
Study            & Year & Method                                          & Split ratio              & WAR & UAR \\
\midrule
Y. Li \textit{et al.} \cite{sota34}           & 2019 & 1-D CNN                                           & 8:2 hold-out                              & 76.66 & 73.64 \\
G. Assunção \textit{et al.} \cite{sota20}              & 2020 & Logistic Model Tree                              & 5-fold CV           & 71.60 & 71.00 \\
Dias Issa \textit{et al.} \cite{sota13}               & 2020 & CNN                                              & 5-fold CV           & 71.61       & -     \\
Mustaqeem \textit{et al.} \cite{sota12}               & 2020 & BiLSTM                                      & 5-fold CV          & 86.00       & 77.00  \\
Mustaqeem and S. Kwon \cite{sota22}      & 2020 & DSCNN & 5-fold CV           & 80.00        & 79.00  \\
T. Tuncer \textit{et al.} \cite{sota17}           & 2021 & SVM                                              & 10-fold CV & 87.43 & \textbf{87.43} \\
\midrule
Our proposed & 2022 & GM-TCNet & 8:2 hold-out & \textbf{90.28}	& \textbf{90.03} \\
Our proposed & 2022 & GM-TCNet & 5-fold CV & \textbf{87.08}	& \textbf{86.91} \\
Our proposed & 2022 & GM-TCNet & 10-fold CV & \textbf{87.64}		& 87.30\\
\bottomrule[1.5pt]
\end{tabular*}
}
\caption{The performances of different approachs on the RAVDESS dataset.}
\label{tab_ravdess}
\end{table*}

\begin{table*}[!h]
\small
\renewcommand\arraystretch{1.5}
\setlength{\tabcolsep}{0.5mm}{
% \begin{tabular}{p{3.3cm}p{0.8cm}p{5.1cm}p{3.3cm}p{1.3cm}p{1.2cm}}
\begin{tabular*}{1\linewidth}{@{\extracolsep{\fill}}lcllcc}
\toprule[1.5pt]
Study            & Year & Method                                          & Split ratio              & WAR & UAR \\
\midrule
T. Özseven \cite{sota26}                                                            & 2019 & SVM                        & 10-fold CV                      & 72.39       & -     \\
N. Hajarolasvadi \textit{et al.} \cite{intro5}                                 & 2019 & 3-D CNN  & 10-fold CV                      & 81.05       & -     \\
F. Daneshfar \textit{et al.} \cite{sota11}                            & 2020 &  Quantum-behaved PSO                     & Leave-OneSpeaker-Out  & 60.79       & -     \\
S. Mekruksavanich \textit{et al.} \cite{sota29}                                              & 2020 &  1-D CNN                  & 10-fold CV                      & 65.83       & -     \\
G. Assunção \cite{sota20}                                                          & 2020 & Logistic Model Tree        & 5-fold CV                       & 70.40 & 68.00 \\
T. Tuncer \textit{et al.} \cite{sota17}                                                      & 2021 & SVM                        & 10-fold CV                      & 84.79 & 83.38 \\
S. Kanwal \textit{et al.} \cite{sota35} & 2021 & Clustering based GA, SVM    & 10-fold CV & 69.80        & -     \\
H. Ibrahim \textit{et al.} \cite{sota36}                  & 2022 & Echo State Networks  & Leave-OneSpeaker-Out   & 68.33 & 64.05 \\
\midrule
Our proposed & 2022 & GM-TCNet & 8:2 hold-out & \textbf{90.63}	& \textbf{91.04} \\
Our proposed & 2022 & GM-TCNet & 5-fold CV &\textbf{84.79}	& \textbf{83.33}\\
Our proposed & 2022 & GM-TCNet & 10-fold CV & \textbf{86.01}	& \textbf{84.40} \\

\bottomrule[1.5pt]
\end{tabular*}
}

\caption{The performances of different approaches on the SAVEE dataset.}
\label{tab_savee}
\end{table*}

\subsubsection{SAVEE}

%On the SAVEE dataset, the proposed method achieves the highest accuracy score of 90.63\%. Among them, Angry, Happy and Disgust have achieved 100\% accuracy. In the previous studies, Tuncer \textit{et al.}\cite{sota17} achieved an accuracy of 84.79\%, which was the best accuracy ever reported. But it is 5.84\% lower than GM-TCNet. \cite{sota29} and our work both use MFCC feature and 1-D convolutional layers, but GM-TCNet beats their method with the 24.8\% higher accuracy.

%Table \ref{tab_savee} shows that GM-TCNet achieves the highest accuracy and the highest accuracy on Fear, Neutral, Sad, and Surprise emotions.

Table \ref{tab_savee} shows that the proposed method achieves the highest accuracy score of 90.63\%, 84.79\% and 86.01\% on the hold-out, 5-fold CV and 10-fold CV respectively on the SAVEE dataset. Although the method proposed by Tuncer \textit{et al.}\cite{sota17} achieved the best-reported accuracy score, 84.79\%, on the 10-fold CV, it is 1.22\% lower than that of GM-TCNet. On the other hand, both Mekruksavanich \textit{et al.}\cite{sota29} and our work use the MFCC feature and 1-D convolutional layers, but GM-TCNet beats the former by achieving 20.18\% higher accuracy.

\subsection{Ablation Experiments}

%As mentioned in Section \ref{proposed algorithm}, we inject the gating mechanism and skip connection into the GM-TCNet architecture. This adoption aims to enhance its discriminative ability and mine potential multi-scale temporal features. Moreover, the Dilated Rate Distribution (DRD) is also adjusted to extract more abundant temporal information based on the greater receptive field. 
%This subsection also gives the comparisons on the performance of GM-TCNet and TCNN based on the same parameter settings to verify the effectiveness of our GM-TCNet. 

\subsubsection{The Gating Level}
%In order to explore the interpretability of each part of the network architecture, the ablation experiment is conducted on the gating level in each GCB. 
This ablation experiment is performed at each GCB gating level, aiming to explore the influence of different gating levels on the performance of GM-TCNet.

\begin{figure*}[t]
	\centering
	\includegraphics[width = 7.2cm]{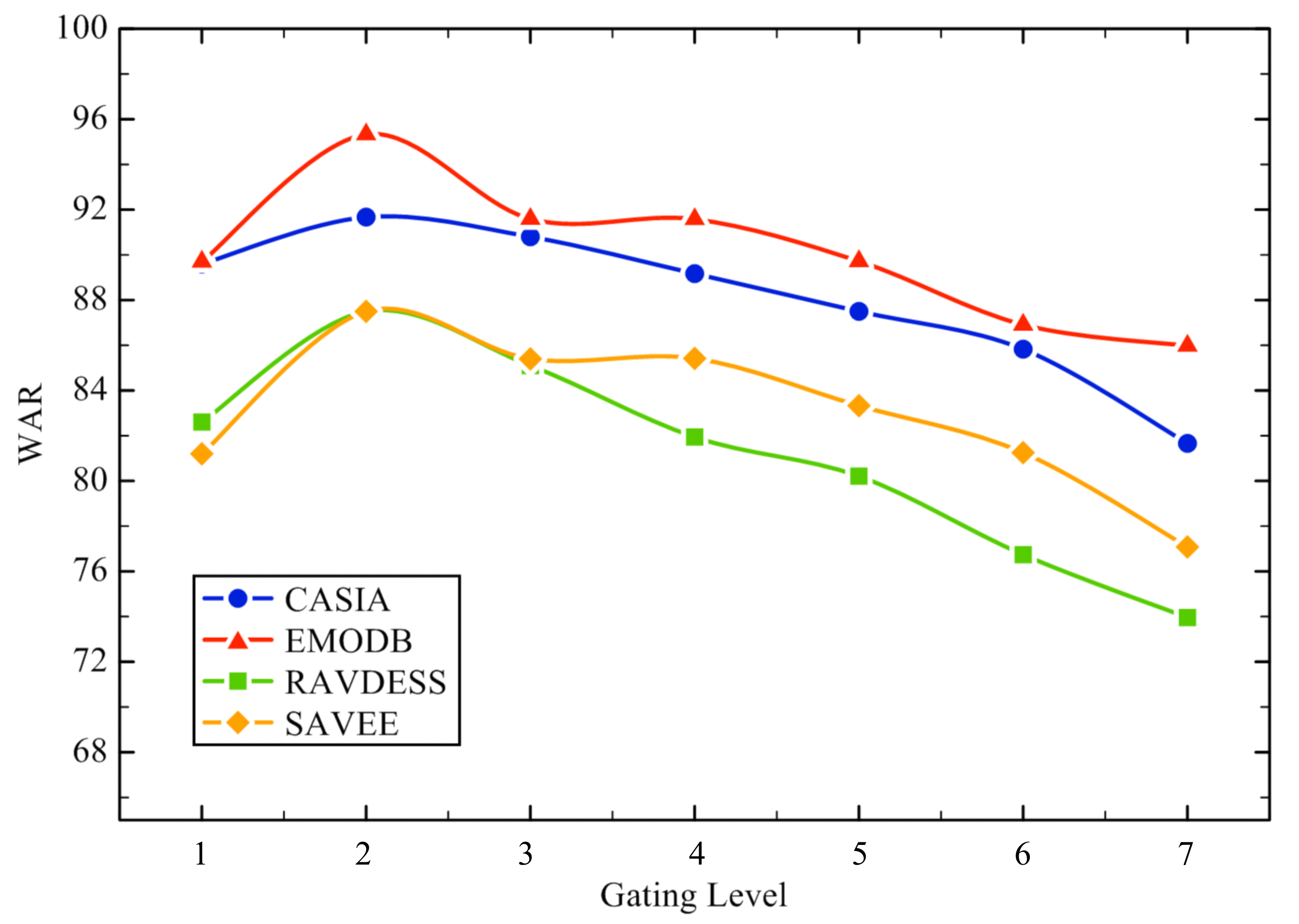}
	\caption{The curves reflect the influence of different gating levels on the accuracy of the model based on 4 datasets.}
	\label{Fig4.10}
\end{figure*}

As shown in Figure \ref{Fig4.10}, the WAR scores are the highest when the gating level is 2 and the other parameters are kept unchanged. When the gating level is larger than 2, the WAR scores decrease. The results show that the two-level gating strategy leads to the best performances on the four datasets. Therefore, the two-level gating can capture key features in the speech signals better than the one-level gating. The reason lies in that the second level gating adds an output gate to control the units' states, providing the network structure with more robust capabilities to filter out irrelevant features. When the gating level is higher than 2, the sigmoid function will attenuate the signals to the degree that the intensity of the necessary information is faded, which results in losing part of the critical feature information in the neural network and hence deteriorating the performance.

Meanwhile, the higher gating level requires more training data to optimize the corresponding parameters. Due to the small sample size in the training data, there should not be too many gating levels in the neural network. Otherwise, the under-fitting problem would occur.

\subsubsection{The Number of GSCB}

\begin{figure*}[t]
	\centering
	\includegraphics[width = 15cm]{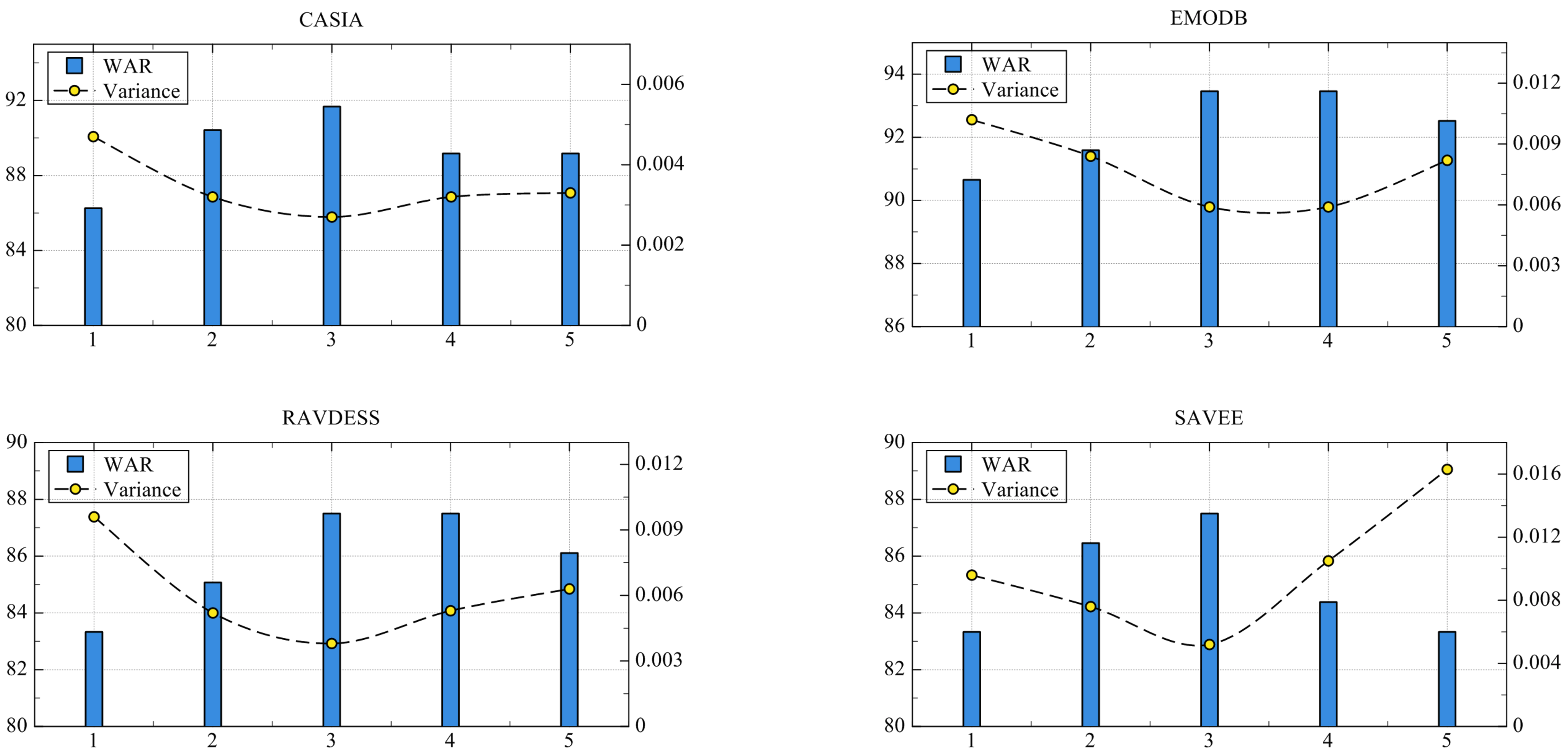}
	\caption{The histograms and the curves respectively reflect the influence of different numbers of GSCB on the accuracy and variance of the model based on 4 datasets.}
	\label{Fig.ensemble}
\end{figure*}

This experiment aims to discover the impact of the number of GSCBs in each gating level on the performance of GM-TCNet. The results in Figure \ref{Fig.ensemble} show that the best GSCB number in each layer is 3, which promises the highest performance on the four datasets. It can effectively enhance the stability and performance of GM-TCNet. The performance deteriorates when the number is too small because of its high variance and low stability. When the number of GSCBs is larger than 3, the variances of models on different datasets increase, and the WAR scores decrease. The larger number of GSCBs would require more parameters to learn, making the model hard to converge. In summary, three GSCBs are the optimal structure for our model, resulting in lower computational cost and better fitting for the real-time application requirement.

 \begin{figure*}[t]
	\centering
	\includegraphics[width = 8.cm]{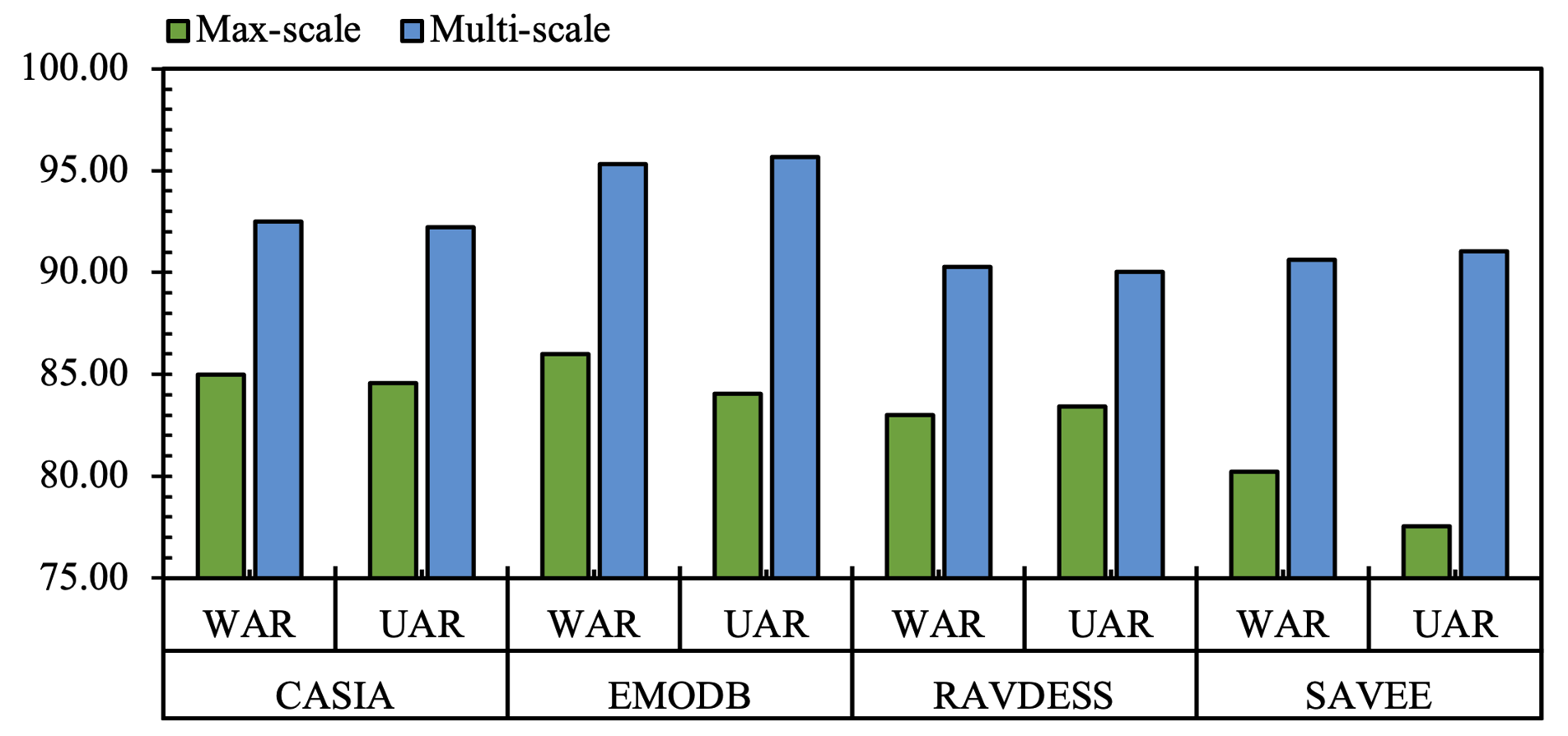}
	\caption{The histograms of WAR and UAR scores obtained by the max-scale and multi-scale receptive field methods on 4 datasets.}
	\label{Fig.multiscale}
\end{figure*}

\subsubsection{The Multi-Scale Temporal Receptive Field}

To figure out the contribution of multi-scale receptive fields, this experiment compares the impact of the max-scale and multi-scale receptive field methods on the performance of GM-TCNet. The max-scale receptive field method only utilizes the output $F_7(x)$ of the last GCB as the input to the LeakyReLU layer, indicating that the receptive field of GTCM is the same as the seventh GCB. The multi-scale receptive field method utilizes the skip connection to sum the outputs $F_i(x)$ from seven GCBs as the input to the LeakyReLU layer. The multi-scale method makes GTCM to obtain multi-scale receptive fields from different GCBs.

The results in Figure \ref{Fig.multiscale} show that the multi-scale method can gain +8.64\% and +9.84\% relative improvement on WAR and UAR compared to the max-scale method. Since the receptive field of a single scale cannot adapt to the changes of different emotions on a time scale, GM-TCNet uses skip connection to combine features from different receptive fields to capture richer multi-scale details. It can effectively enhance the capability of dynamic perception emotion of GM-TCNet.

\subsubsection{The Distribution of Dilated Rates}
\label{DRD}

\begin{figure*}[t]
	\centering
	\includegraphics[width = 15cm]{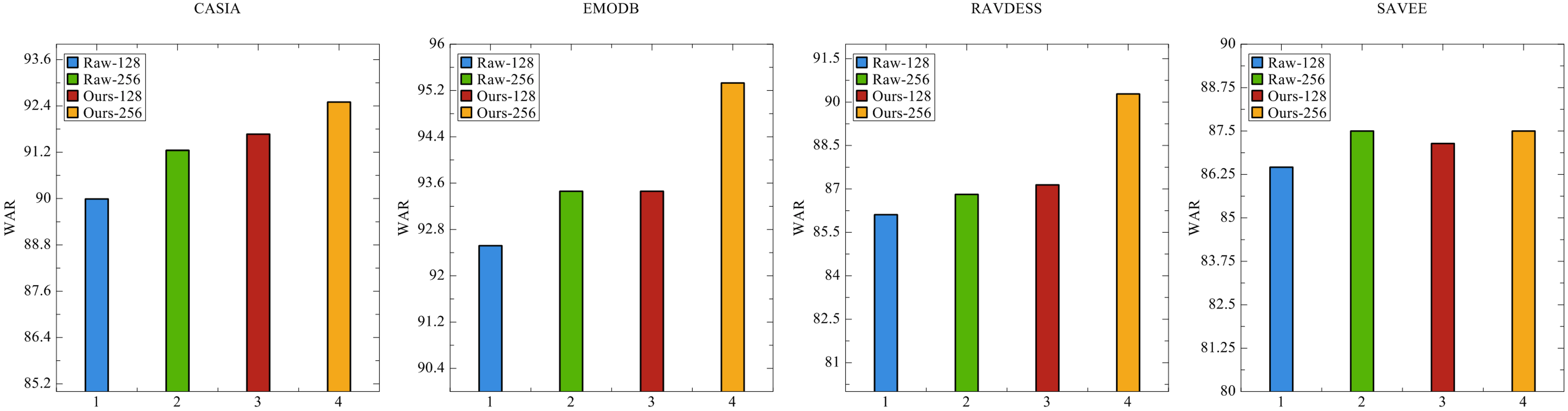}
	\caption{The WAR score histograms of different Dilated Rate Distributions on 4 datasets.}
	\label{Fig4.11}
\end{figure*}

The dilated causal convolutional layer is one of the widely-used basic structures in GSCB. Notably, the dilated rate setting can strongly affect the size of the receptive field and the extraction of high-level features. Therefore, we set diverse values according to the corresponding levels of GCBs. This subsection conducts experiments to explore the influence of different Dilated Rate Distribution (DRD) on performance. Since the feature lengths of the four datasets range within [128, 256] after feature extraction, we set the receptive field sizes in the range of [128, 256] in this experiment. In Figure \ref{Fig4.11}, "Raw-128/256" means to use TCNN's original DRD on GM-TCNet, that is, 7/8 layers are deployed to generate 128/256 receptive fields, respectively. Similarly, "Ours-128/256" means to use the proposed new DRD on GM-TCNet in which 6/7 layers are used to get 128/256 receptive fields.

\begin{table}[htbp]
\centering
\small
\renewcommand{\arraystretch}{1.5}
% \begin{tabular*}{cccccccc}
\begin{tabular*}{1\linewidth}{@{\extracolsep{\fill}}ccccc}
\toprule[1.5pt]
Dataset & Raw-128 & Raw-256 & Ours-128 & Ours-256 \\
\midrule
CASIA   & 89.99         & 91.25          & 90.42 & \textbf{92.50} \\
EMODB   & 92.52         & 93.46          & 93.46 & \textbf{95.33} \\
RAVDESS & 86.11         & 86.81          & 87.14 & \textbf{90.28} \\
SAVEE   & 86.46         & 87.50          & 86.91 & \textbf{89.60} \\\midrule
\textit{Parameters}      & 0.261M         & 0.298M     & 0.224M      & 0.261M \\
\bottomrule[1.5pt]
\end{tabular*}
\caption{The WAR scores of different DRD method used in GM-TCNet.}
\label{Tab4.2}
\end{table}

Table \ref{Tab4.2} shows that our DRD leads to the best results on four datasets compared with the original DRD. When the number of layers is the same, our DRD obtains a larger receptive field than the original DRD. It is beneficial to capture the global features in the time domain. The experiment results show that the best WAR scores of the proposed method is +2.92\% higher than the original DRD on the four datasets on average, indicating that the adjusted receptive field offers sentimental information on more temporal scales and higher capability in utilizing the temporal information.

When the size of the receptive field is unchanged, the original DRD requires one more layer than the proposed method. Table \ref{Tab4.2} shows that a more complex network structure leads to higher training cost with worse performance. When the receptive field is set to 256, our DRD beats the original one with +1.94\% higher in WAR on the four datasets on average. Therefore, the proposed DRD fits the SER applications better with higher generalization ability, which reveals that different speech corpus might need much longer memory.

\subsection{Interpretability Analysis of GM-TCNet and High-Level Features}
  
The input to GM-TCNet is the MFCC feature that is composed of 39-D cepstral coefficients. We can obtain the coefficients after the discrete cosine transform processes the speech frequency spectrum. The frequency spectrum of the speech signal can be regarded as the superposition of the low-frequency envelope and the high-frequency details. Specifically, the low-frequency component of the cepstrum is the envelope of the spectrum. The envelope is a smooth curve connecting the formant points. The spectrum contains the high-frequency components of the cepstrum, which are the amplitude near the formant point.

Because the discrete cosine transformation has the characteristic of energy concentration, most of the energy information of the signal data falls in the low-frequency area after the discrete cosine transformation. The MFCC features contain many envelopes and lack explicit physical meaning. Whereas, GM-TCNet can capture high-level features with emotion discrimination in both the time and frequency domains. The analysis is given from three perspectives.

\subsubsection{Extraction Process of High-Level Features}

\begin{figure*}[t]
	\centering
	\includegraphics[width = 14cm]{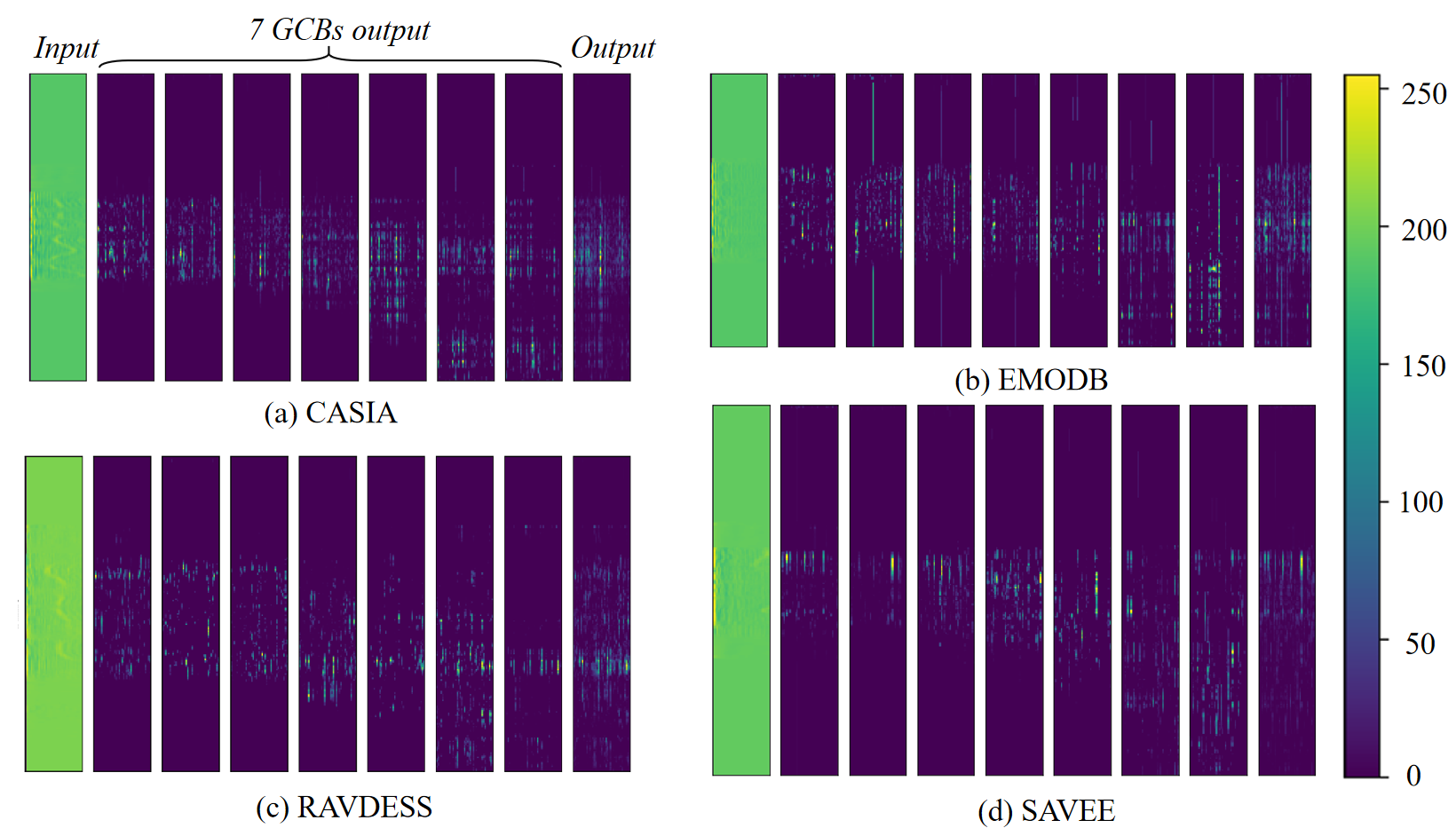}
	\caption{The visualized results of the MFCC feature. From left to right, the nine bars are: input, the outputs of seven GCBs, and the output of GTCM.}
	\label{Fig4.14}
\end{figure*}

The extracted high-level features from the hidden layers of the GM-TCNet are illustrated in Figure \ref{Fig4.14}. The feature information extracted by the output of different GCB is diverse in time and frequency domains. The skip connection is employed to fuse the high-level features captured by different GCBs with the diverse receptive fields for the time domain. When the network deepens, the receptive field of the GCB is enlarged exponentially. Usually, the small receptive field perceives more detailed information from the local features, while the large receptive field perceives global information to ensure a longer range of sentimental dependency. Therefore, the features extracted by the low-level GCB contain detailed information and the high-level GCB expands the receptive field through a larger dilated rate to perceive global information. This structure promises high diversity among the features extracted from multi-scale temporal receptive fields to maintain the robust discriminative ability.

As shown in Figure \ref{Fig4.14}, GM-TCNet offers similar feature extraction capabilities on the four datasets. In the visualization of each dataset, the first and last feature maps represent the MFCC and the output of the GTCM, respectively. The hidden feature map shows the output of each GCB in the order from left to right. The values of these feature maps are normalized in the range of [0, 255], and then are visualized in Figure \ref{Fig4.14}. It can be observed that the feature maps extracted from GCBs with larger receptive fields tend to show more information in the time domain. Moreover, the last feature map shows that the skip connection can make GTCM capture multi-scale feature and enhance the capability of dynamic perception emotion of GM-TCNet.

\subsubsection{Comparisons Between the High-Level Feature and MFCC Feature}

\begin{figure*}[t]
	\centering
	\includegraphics[width = 7.5cm]{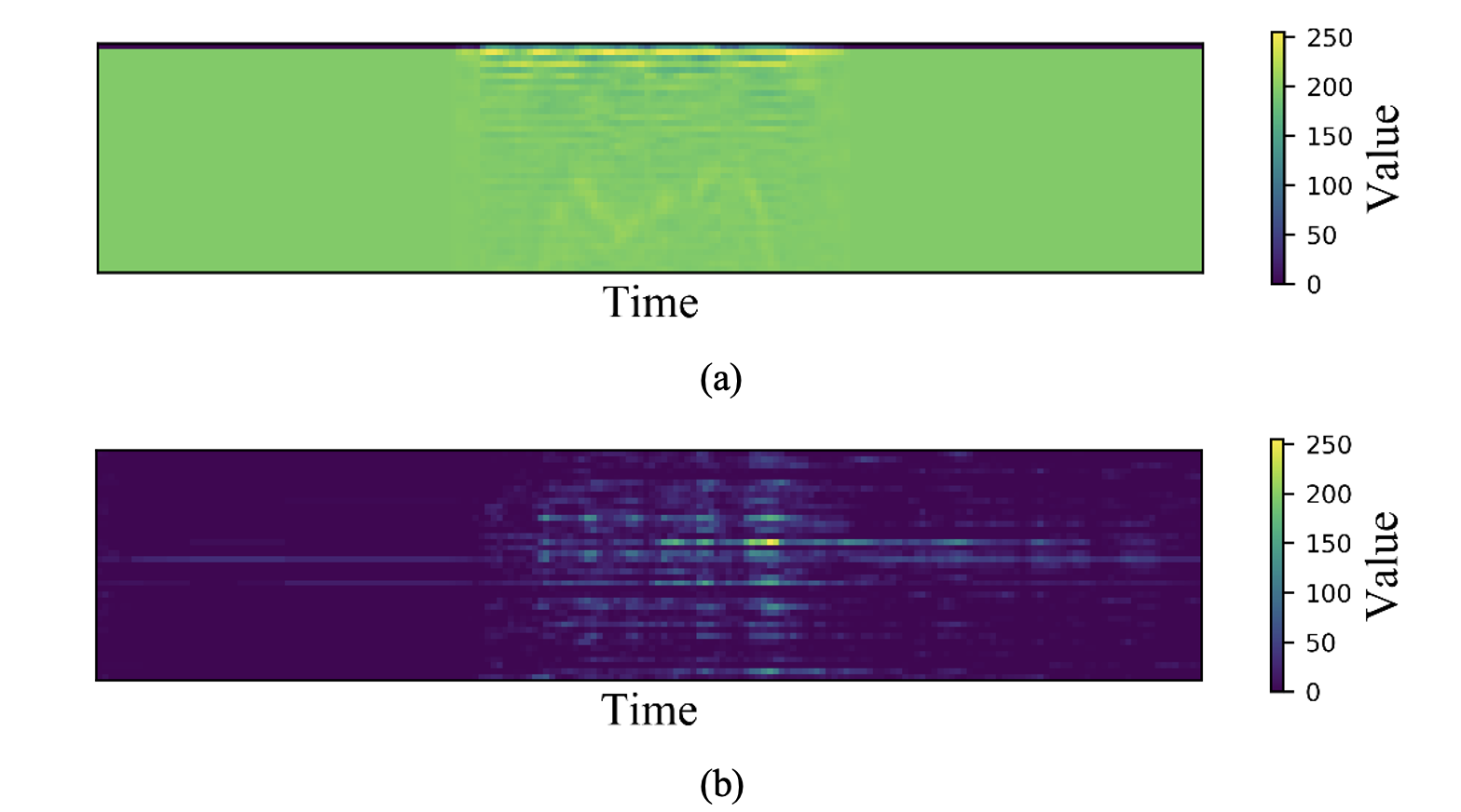}
	\caption{MFCC feature (top) and high-level feature (bottom) visualizations without global 1-D average pooling}
	\label{Fig4.12}
\end{figure*}

\begin{figure*}[t]
	\centering
	\includegraphics[scale = 0.32]{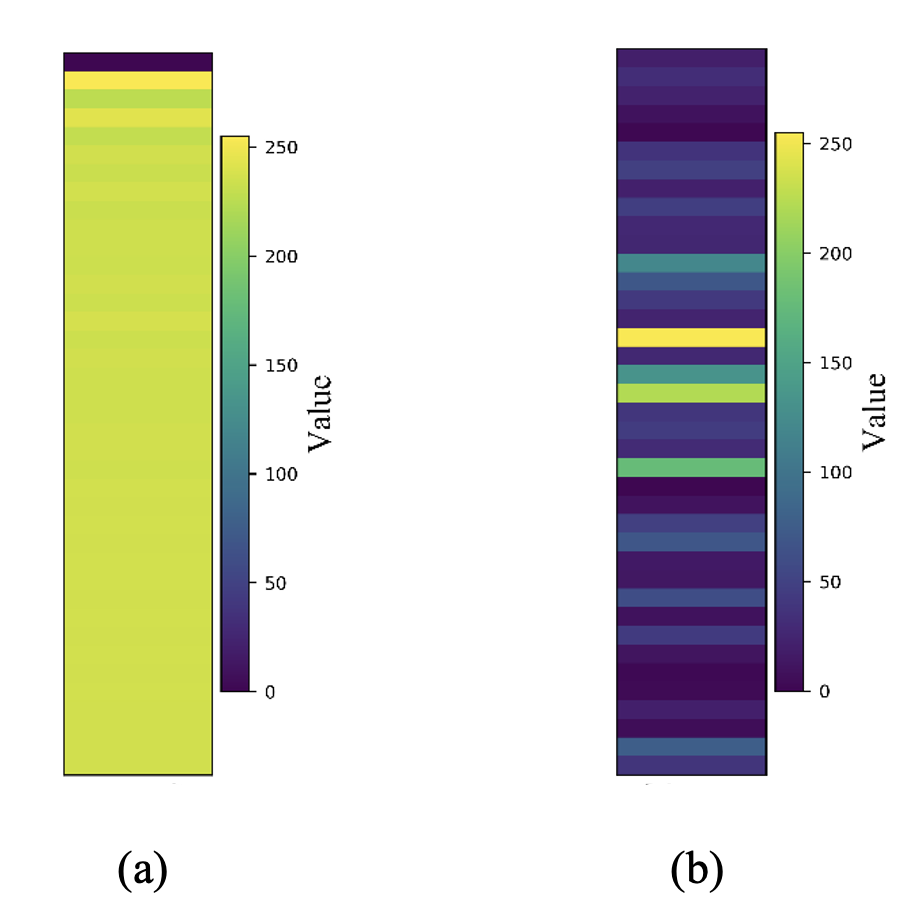}
	\caption{MFCC feature (left) and high-level feature (right) visualization with global 1-D average pooling}
	\label{Fig4.13}
\end{figure*}

In this section we compare the MFCC features with the high-level features extracted by GTCM in the time domain and the frequency domain. Figure \ref{Fig4.12} shows that the initial MFCC features contain a large amount of redundant information. In contrast, the high-level features are the outputs of the GTCM, which only retain a small quantity of critical information in the speech signals. In the MFCC features, except for the logarithmic energy coefficient of the first dimension, the diversity of the cepstral coefficients in other dimensions is low. However, the high-level feature of GM-TCNet is significantly different in each dimension to produce more discriminative representation for the SER task. Moreover, Figure \ref{Fig4.13} confirms that GM-TCNet can capture characteristic information in the time domain and frequency domain simultaneously.

\subsubsection{Analysis of High-Level Features among Different Emotions}

\begin{figure*}[t]
	\centering
	\includegraphics[width = 9.5cm]{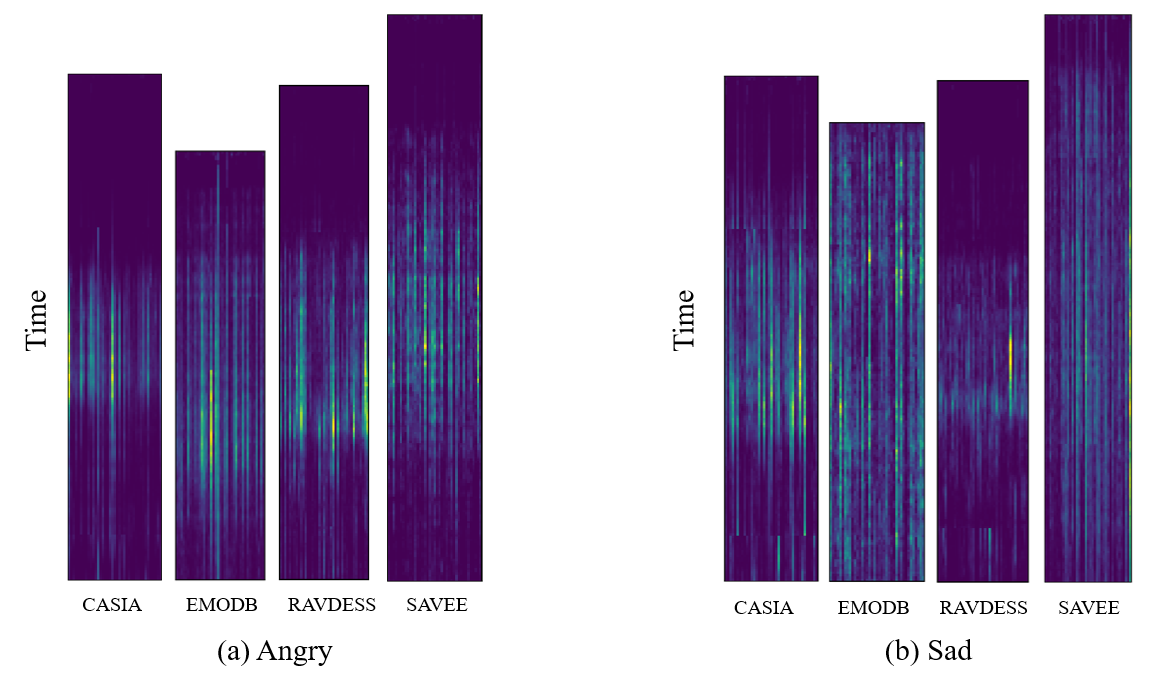}
	\caption{The comparison of high-level feature visualization between Angry and Sad on four datasets}
	\label{Fig4.15}
\end{figure*}

Arousal and valence are two independent dimensions of the continuous emotional model \cite{arousal_valence}. Arousal represents how excited or indifferent the emotion is, and valence represents how positive or negative the emotion is. Almost all human emotions can be represented by the 2-D space formed by arousal and valence. As shown in Figure \ref{Fig4.15}, for the high-arousal speech signals, such as Angry, Happy, and Surprise \cite{intr_1}, the extracted high-level features show the high intensive distributions. It means a larger amount of information tends to burst out in a short time \cite{emo_fluct}. In contrast, for the low-arousal speech signals, such as Sad, Calm, and Boredom \cite{intr_2}, the high-level features give the more balanced distributions. The results indicate that the GTCM provides more robust and discriminative representations for the emotion classifier to support superior performance.

In order to get the insights into the contributions of the high-level features, the information entropy is introduced to evaluate the contributions of the high-level features in various emotions. The information entropy is a quantitative assessment of the information expressed by the image, which reflects how much information there is in the image~\cite{entropy}. We calculate entropy values for the normalized high-level feature maps from GTCM.

Specifically, it is assumed that the size of the high-level feature map $M$ is $W\times H$, and the size of $M'$ after zero padding is $(W+2)\times (H+2)$. The average value of the $3\times 3$ neighborhood $G(x, y)$ corresponding to the pixel point $(x, y)$ is defined in Eq.(\ref{E_entropy1}), where $1\leq x\leq W, 1\leq y\leq H$. In particular, $K$ represents the number of non-padding elements in the region of $3\times 3$ neighborhood. For every point, a 2-tuple $(m,n)$ is introduced to show that the pixel point $(x,y)$ has the property of $M(x,y)=m$ and $G(x,y)=n$. In addition, $r_{mn}$ represents the frequency of the 2-tuple $(m,n)$. Then the joint probability density $P_{mn}$ is obtained by Eq.(\ref{E_entropy2}), and the 2-D entropy $E$ is calculated by Eq.(\ref{E_entropy3}).

\begin{gather}
G(x, y)=\dfrac{(\sum\limits_{i=-1}^{1} \sum\limits_{j=-1}^{1} M'(x+i, y+j)) - M(x,y)}{K-1} \label{E_entropy1}\\
P_{mn} = \dfrac{r_{mn}}{W\times H}\label{E_entropy2}\\
E=-\sum_{m=0}^{255} \sum_{n=0}^{255} P_{mn} \log _{2} P_{mn} \label{E_entropy3}
\end{gather}

As shown in Table \ref{Tab4.1}, the entropy values of Angry, Happy, and Surprise emotions are higher than those of Sad, Calm, and Boredom emotions. The former three can be classified as excited in arousal, and the last three can be classified as indifferent. This is in accordance with the mapping of the diverse emotion groups onto the arousal in \cite{ser_sim1, ser_sim2}. The results indicate that GM-TCNet can effectively distinguish the emotions in binary arousal.

\begin{table}[htbp]
\centering
\small
\linespread{1} 
\renewcommand{\arraystretch}{1.5}
% \begin{tabular}{cccccccc}
\begin{tabular*}{1\linewidth}{@{\extracolsep{\fill}}cccccccc}
\toprule[1.5pt]
\multicolumn{2}{c}{CASIA} & \multicolumn{2}{c}{EMODB} & \multicolumn{2}{c}{RAVDESS} & \multicolumn{2}{c}{SAVEE} \\
\midrule
angry       & 14.1182     & angry       & 13.0694     & angry        & 14.1036      & angry       & 14.5365     \\
happy       & 13.9951     & happy       & 13.2558     & happy        & 14.2427       & happy       & 13.5133     \\
surprise       & 14.0909     & surprise       & -     & surprise        & 13.9349      & surprise       & 14.7997     \\
\specialrule{0em}{0.6pt}{0.6pt}
\hline
\specialrule{0em}{0.6pt}{0.6pt}
sad         & 13.4459     & sad         & 12.8941     & sad          & 13.7379      & sad         & 13.3498     \\
calm     & -           & calm     & -     & calm      & 13.8056      & calm     & -    \\
boredom     & -           & boredom     & 13.0693     & boredom      & -      & boredom     & -    \\
\bottomrule
\end{tabular*}
\caption{2-D entropy of different emotions.}
\label{Tab4.1}
\end{table}
\begin{figure*}[t]
	\centering
	\includegraphics[width = 9.5cm]{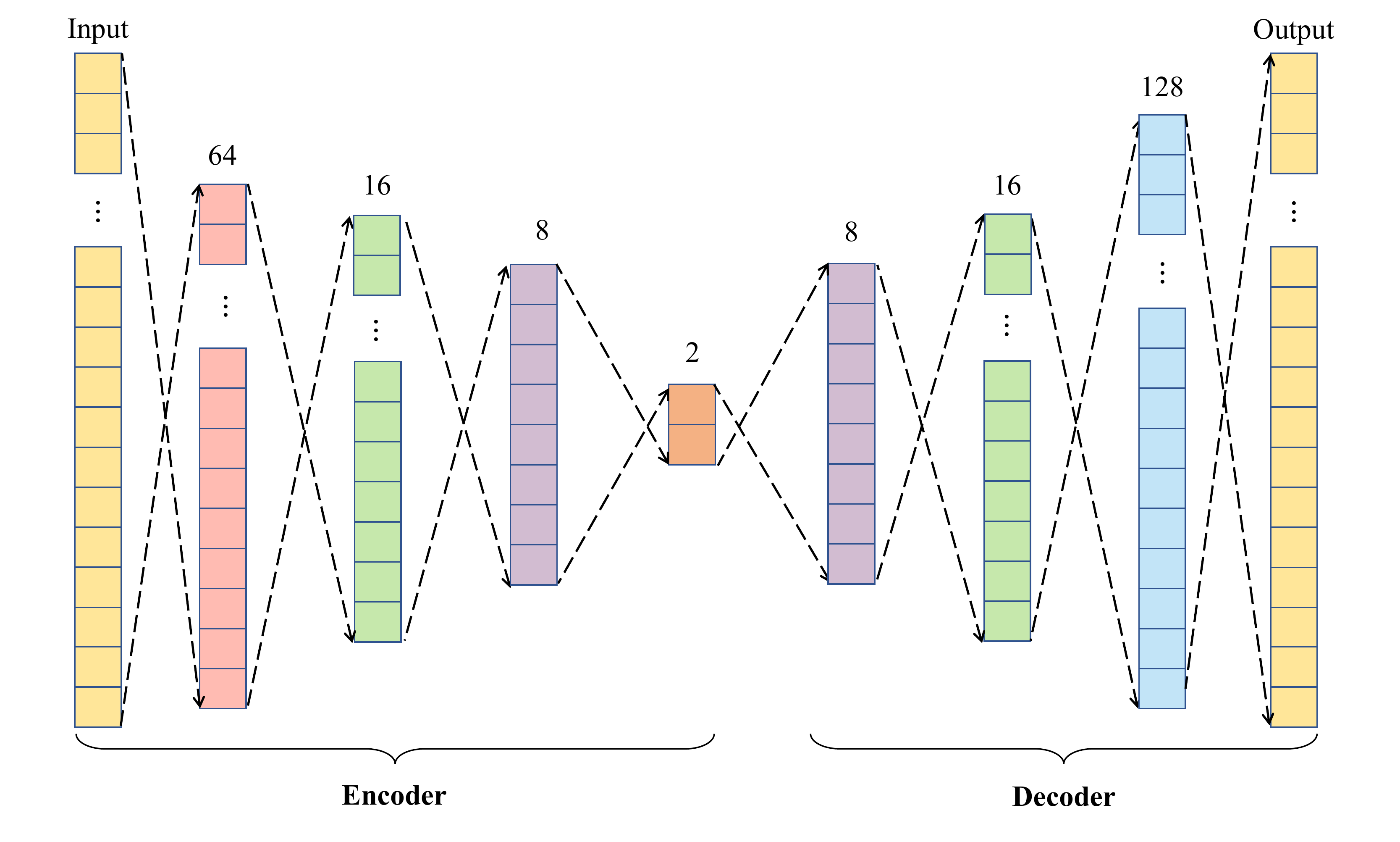}
	\caption{The network structure of autoencoder.}
	\label{Fig4.16}
\end{figure*}
Furthermore, an autoencoder (AE) is deployed to project the high-level features into low dimensional feature space to explore the difference in high-level features across diverse emotion classes. As shown in Figure \ref{Fig4.16}, AE is a particular type of neural network composed of an encoder and decoder. And it is trained on the encoded data, and outputs a recreation of that data. In this experiment, the encoder consists of four FC layers with 64, 16, 8 and 2 neurons. The encoder can mine the low dimension representation of high-level features. The decoder consists of four FC layers, and the first three layers contain 8, 16 and 128 neurons. While the last layer includes the same number as the compressed length of the high-level feature extracted by GTCM. In this way, the decoder ensures a small deviation between the reconstructed features and high-level features. 

\begin{figure*}[t]
	\centering
	\includegraphics[width = 14cm]{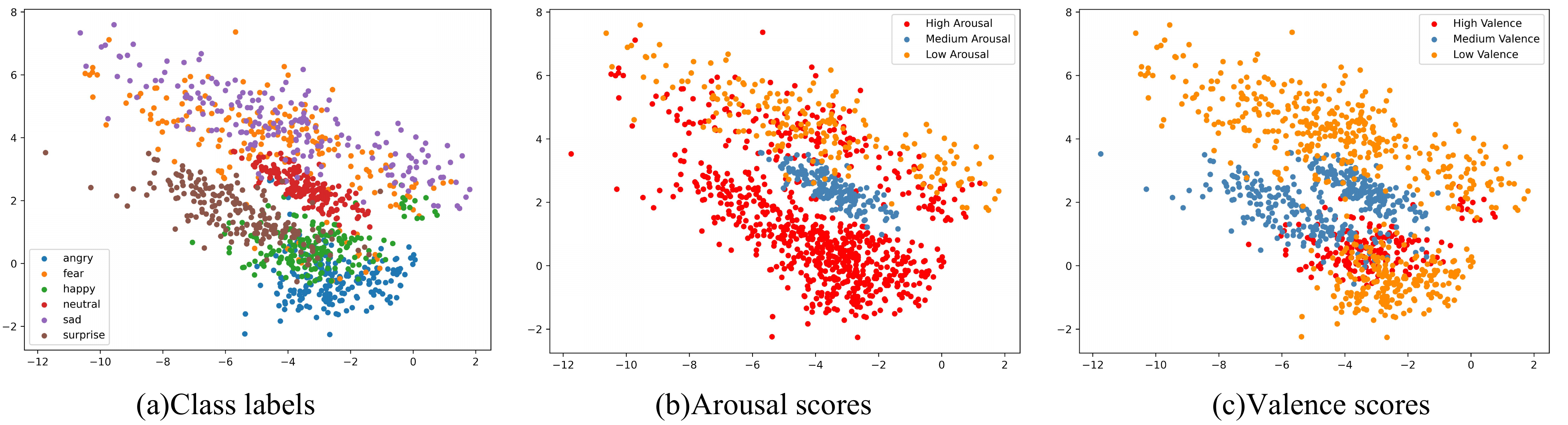}
	\caption{High-level features visualizations of the AE representations for CASIA.}
	\label{Fig4.17}
\end{figure*}
\begin{figure*}[t]
	\centering
	\includegraphics[width = 14cm]{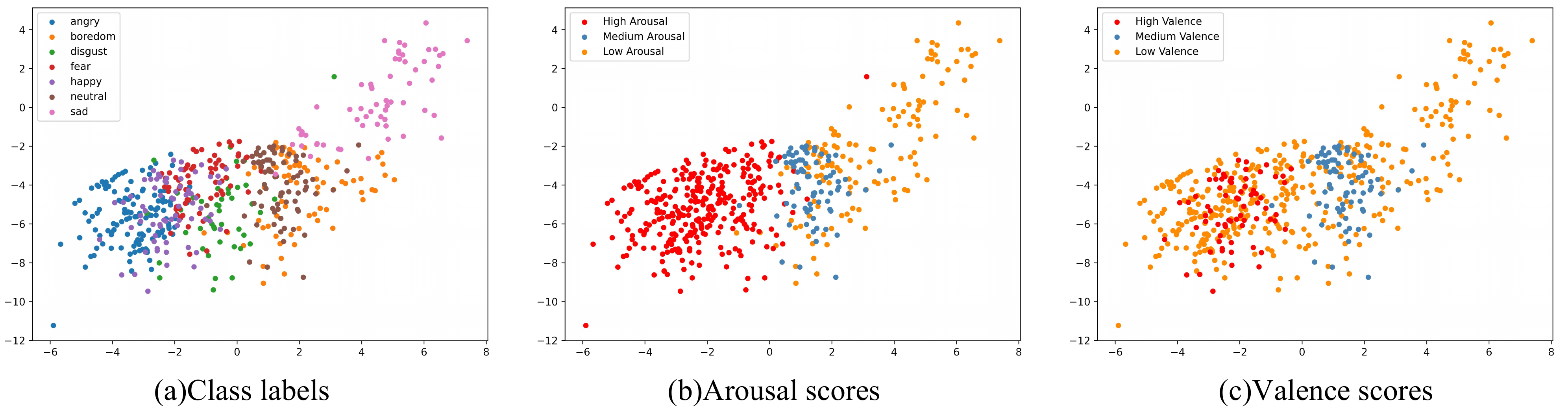}
	\caption{High-level features visualizations of the AE representations for EMODB.}
	\label{Fig4.18}
\end{figure*}
\begin{figure*}[t]
	\centering
	\includegraphics[width = 14cm]{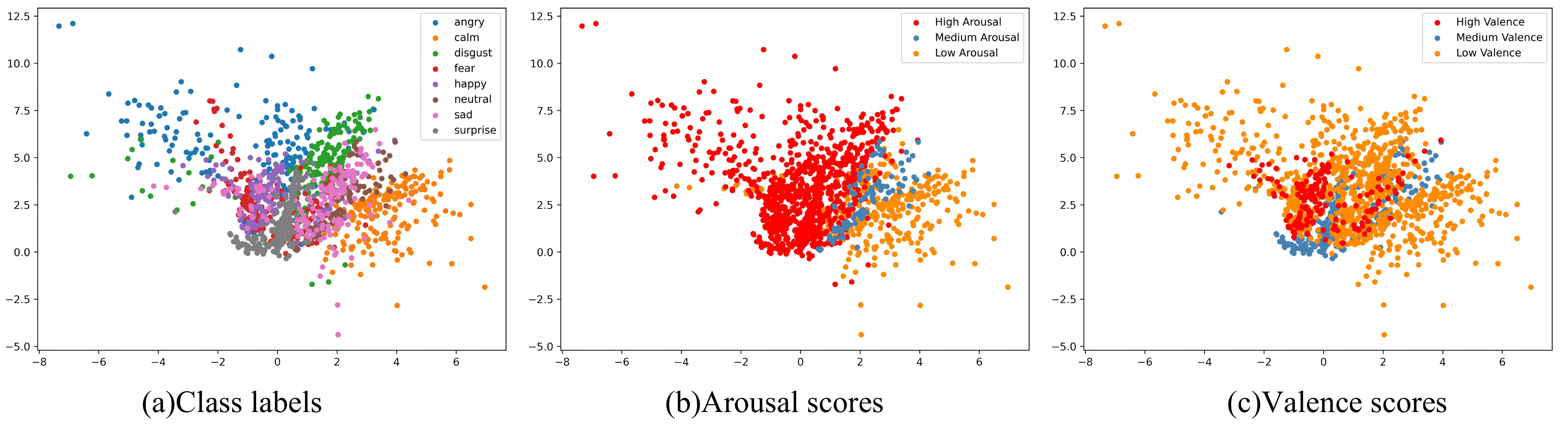}
	\caption{High-level features visualizations of the AE representations for RAVDESS.}
	\label{Fig4.19}
\end{figure*}
\begin{figure*}[t]
	\centering
	\includegraphics[width = 14cm]{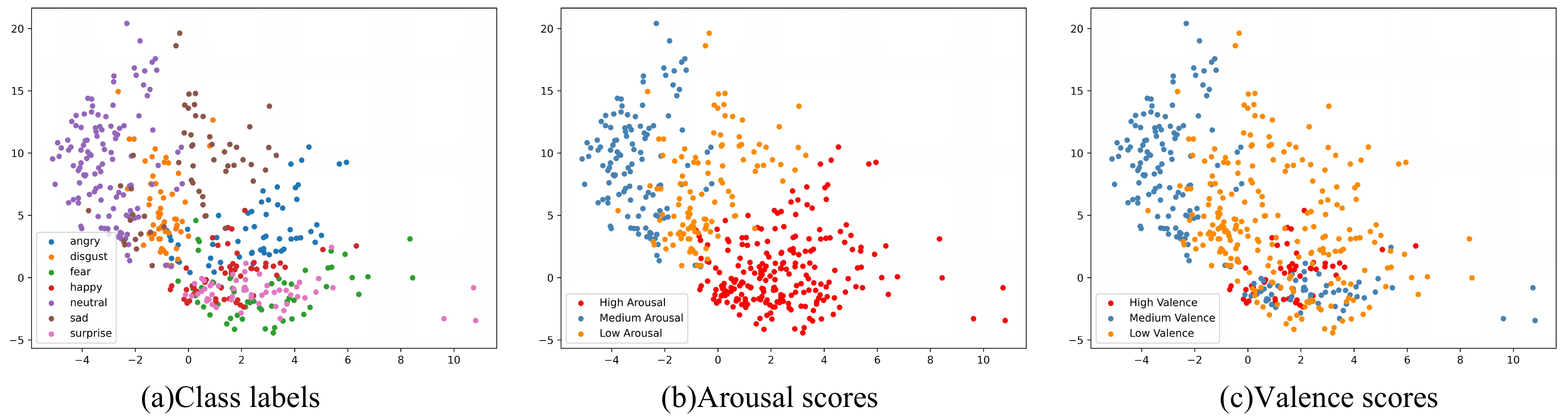}
	\caption{High-level features visualizations of the AE representations for SAVEE.}
	\label{Fig4.20}
\end{figure*}

Figure \ref{Fig4.17}-\ref{Fig4.20} show the 2D projections generated by AE. These figures depict that the samples in each cluster mainly belong to the same emotion, so there are large margins among the boundaries of different classes. Figure \ref{Fig4.17}(a) shows that the features can well split the Angry, Happy, Neutral, and Surprise emotions since the corresponding clusters are of high purity. While the clusters representing the Fear and Sad emotions are mixed with samples from other emotions, leading to the relatively high error rates in the two emotions. Besides, the scatter plots for the arousal scores in Figure \ref{Fig4.17}(b) and the valence scores in Figure \ref{Fig4.17}(c) show that the GM-TCNet is more discriminative for the arousal scores than for the valence scores. Specifically, the distributions of the arousal scores in Figure \ref{Fig4.17}(b) show the gradual change trend. In contrast, the distributions of valence scores present heavy overlappings caused by the high variance in the cluster of low valence scores. The samples with low valence scores get high error rates. These observations align with the findings in \cite{ser_sim1}.

Similar conclusions can be drawn from Figure \ref{Fig4.18} and Figure \ref{Fig4.19}. In the EMODB dataset, the Happy emotion forms a cluster that largely overlaps with the Angry emotion, which matches the results in Figure \ref{EMODB_matrix}. A similar observation is found in Figure \ref{Fig4.19}(a) on the RAVDESS dataset. Additionally, Figure \ref{Fig4.18} and Figure \ref{Fig4.19} also show that the clusters of arousal emotion are of higher purity than those of the valence emotions, confirming that GM-TCNet provides a higher discriminative ability for the arousal emotions than for the valence emotions. 

It is interesting to find that the patterns shown in Figure \ref{Fig4.20} are pretty different from those on other datasets. The difference reflects that the distribution of the arousal emotions on the SAVEE dataset is of high diversity from those of other datasets. The arousal score distribution in Figure \ref{Fig4.20}(b) does not show the continuous distribution from the high-arousal to low-arousal. However, Figure \ref{Fig4.20}(b) and Figure \ref{Fig4.20}(c) still exhibit that the three arousal levels change gradually, implying that GM-TCNet can separate the arousal emotions better than the valence emotions.

\section{Conclusions}
\label{Conclusions}

This paper discusses the SER task by proposing a novel GM-TCNet approach based on the dilated causal convolution and gating mechanism. GM-TCNet is designed to explore causal relationships and long-term dependencies among different emotions. A novel emotional causality representation learning component is designed to capture the dynamics of emotion across the time domain. It also has a strong ability to build a reliable long-term sentimental dependency. It is the first attempt at applying the causality learning method to SER to the best of our knowledge. The experimental results confirm that mining emotional causality in speech is of great significance for the SER task. 

In the consideration that the human speech expression is not single-scale but multi-scale in nature, GM-TCNet uses the skip connection among all Gated Convolution Blocks. It provides our network structure with a multi-scale temporal receptive field, enhancing the model's speech emotion perception. Moreover, a new dilated rate distribution of blocks is designed to obtain a larger receptive field, so as to better fit the SER applications with higher generalization ability. Compared with the widely deployed methods for SER that used multi-modal features, we believe that the information embedded in a single type of feature can support high discriminative ability given an effective mining scheme. Therefore, this study only deploys the standard MFCC feature to extract high-level features by our GM-TCNet. 

Experiment results verify that GM-TCNet successfully captures the high-level features of speech in the time domain. Compared with other studies, it obtains the highest accuracies on the four commonly used datasets in most cases compared to SOTA techniques.

However, since the speech datasets used in this study are audio files with short duration, the performance of GM-TCNet in the real-world applications still needs to be further tested. Therefore, our future work will focus on enhancing the generalization ability of the long-duration audio data. At the same time, more efforts will be made to explore other types of features, especially those in the time domain.
%% The Appendices part is started with the command \appendix;
%% appendix sections are then done as normal sections
% \appendix

% \section{}
% \label{}

%% If you have bibdataset file and want bibtex to generate the
%% bibitems, please use
%%

%%  \bibliography{<your bibdataset>}

\section*{Acknowledgement}
This work is supported by the National Natural Science Foundation of China (No. 61772023), National Key Research and Development Program of China (No. 2019QY1803), and Fujian Science and Technology Plan Industry-University-Research Cooperation Project (No.2021H6015). The algorithm production is supported by the AutoDL.com platform.

{\bibliographystyle{elsarticle-num}%
\bibliography{reference}}

%% else use the following coding to input the bibitems directly in the
%% TeX file.

%\begin{thebibliography}{00}

% \bibitem[Author(year)]{label}
% Text of bibliographic item

%\end{thebibliography}
\end{document}